\documentclass[12pt,a4paper]{article}
\usepackage{jheppub}
\usepackage[FIGTOPCAP]{subfigure}
\usepackage{amsmath}
\usepackage{amsfonts}
\usepackage{amssymb}
\usepackage{tensor}
\usepackage{booktabs}
\usepackage{empheq}
\usepackage{amsmath}
\usepackage{amssymb}
\usepackage{amsthm}
\usepackage{comment}
\usepackage{epigraph}
\usepackage{psfrag}
\usepackage{graphicx}
\usepackage{color}
\usepackage{bbm}
\usepackage{color}

\newcommand{\lfp}{\text{\tiny LCL}}

\newcommand{\exclude}[1]{}

\newcommand{\beq}{\begin{equation}}
\newcommand{\eeq}{\end{equation}}
\newcommand{\bea}{\begin{eqnarray}}
\newcommand{\eea}{\end{eqnarray}}
\long\def\/*#1*/{}

\newcommand{\junk}[1]{}

\title{Holographic Lifshitz flows}

\author[a,b,c]{Matteo Baggioli,}
\author[d]{Oriol Pujol{\`a}s,}
\author[a,b,c]{Xin-Meng Wu}
\affiliation[a]{School of Physics and Astronomy, Shanghai Jiao Tong University, Shanghai 200240, China}
\affiliation[b]{Wilczek Quantum Center, School of Physics and Astronomy, Shanghai Jiao Tong University, Shanghai 200240, China}
\affiliation[c]{Shanghai Research Center for Quantum Sciences, Shanghai 201315, China}
\affiliation[d]{Institut de F\'isica d'Altes Energies (IFAE), The Barcelona Institute of Science and Technology (BIST), 
Campus UAB, 08193 Bellaterra, Barcelona}

\emailAdd{b.matteo@sjtu.edu.cn}
\emailAdd{pujolas@ifae.es}
\emailAdd{xinmeng.wu@sjtu.edu.cn}

\abstract{Without Lorentz symmetry, generic fixed points of the renormalization group (RG) are labelled by their dynamical (or `Lifshitz')
exponent $z$. Hence, a rich variety of possible RG flows arises. 
The first example is already given by the standard non-relativistic limit, which can be viewed as the flow from a $z=1$ UV fixed point to a $z=2$ IR fixed point. 
In strongly coupled theories, there are good arguments suggesting that Lorentz invariance can emerge dynamically in the IR from a Lorentz violating UV. 
In this work, we perform a generic study of fixed points and the possible RG flows among them in 
a minimal bottom-up holographic model without Lorentz invariance, aiming to shed light on the possible options and the related phenomenology. We find: 
i) A minor generalization of previous models involving a massive vector field with allowed self-couplings leads to a much more efficient emergence of Lorentz invariance than in the previous attempts. Moreover, we find that generically the larger is the UV dynamical exponent $z_{UV}$ the faster is the recovery of Lorentz symmetry in the IR.
ii) We construct explicitly a holographic model with a line of fixed points, realizing different Lifshitz scaling along the line. 
iii) We also confirm the monotonicity of a recently proposed a-function along all our Lorentz violating RG flows.}

\begin{document}

\maketitle
\color{black}

\section{Introduction}
\epigraph{\textit{Meglio tardi che mai.}}{}
The symmetries of physical systems depend on the energy scale of interest, and Lorentz invariance (LI) is no exception. Numerous experimental tests of LI \cite{Kostelecky:2008ts} confirm that it is realized to exquisite precision at high energies. LI seems to be  a fundamental symmetry of nature, that is, of `ultraviolet' (UV) physics. However, LI is often not manifest in the `infrared' (IR). 
In the eyes of quantum field theory, it is natural to see this as a renormalization group (RG) flow.

Fixed (or critical) points refer to field theories that are invariant under a scaling symmetry. A generic scaling transformation that is compatible with rotations but not boosts (and hence not Lorentz invariant) is of the form
\begin{equation}\label{ll}
t\rightarrow \lambda^z\, t,\quad x^i\rightarrow \lambda\,x^i\,,
\end{equation}
with $z$ the `Lifshitz' (or `dynamical') exponent. LI fixed points must have $z=1$ but without LI this is a free parameter that characterizes the IR dynamics. 
In general, the RG flow from the UV to the IR can connect fixed points with different Lifshitz scalings, $z_{UV}$ and $z_{IR}$.

A basic example is realized by the massive free particle. The dispersion relation `flows' from $E\simeq c \,p$ at $E\gg m$ to $E=p^2/2m$ at $E\ll m$ (with $c$, $p$ and $m$ the speed of light, the particle momentum and mass respectively). This is the usual non-relativistic limit induced by particle number, and it corresponds to a flow from a $z_{UV}=1$ to $z_{IR}=2$. The emergence of $z_{IR}=2$ scaling is key for the dynamics too -- indeed it plays a central role in various effective field theories for non-relativistic bound states, see \textit{e.g.} \cite{Georgi:1990um,Luke:1992cs,Brambilla:2004jw,Eby:2014fya,Braaten:2016kzc,Namjoo:2017nia}. More examples of IR dynamics with nontrivial dynamical exponents, $z_{IR}\neq1$, arise \textit{e.g.} in multi-critical points \cite{PhysRevLett.61.2376}, self-ordering dynamics \cite{Bray:1994zz,Berges:2010ez,PineiroOrioli:2015cpb,Moore:2015adu,Mukhopadhyay:2020gwc,Mukhopadhyay:2020xmy,Mukhopadhyay:2024wii}, several quantum critical points \cite{PhysRevB.23.4615,PhysRevLett.35.1678,PhysRevB.64.195109,PhysRevLett.88.217204} or in systems involving spontaneous symmetry breaking at finite density, which can lead to Goldstone bosons with non-linear dispersion relations in some cases \cite{Nielsen:1975hm,Watanabe:2012hr,Hidaka:2012ym}.

The `opposite' type of flow, from $z_{UV}=2$ to $z_{IR}=1$ is also possible in systems at finite density. This occurs for example in crystalline solids, in the mechanical (or phononic) sector.
The `UV theory', consisting of an array of non-relativistic massive particles ($z_{UV}=2$) coupled to near neighbours via Hooke's law, indeed flows in the IR to a theory of gapless phonons ($z_{IR}=1$). This type of flow is also realized in the electronic sector in Dirac and Weyl materials such as graphene \cite{geim2009graphene} or Weyl semimetals \cite{yan2017topological}. As is well known, close to charge neutrality, the quasi-electron excitations are well described by LI fermions (nevertheless, with an emergent lightcone speed different from $c$).

Finally, flows to $z_{IR}=1$ are of key importance for Lorentz violating completions of gravity, such as the Ho\v rava-Lifshitz model of quantum gravity \cite{Horava:2009uw,Blas:2009qj} where $z_{UV}>1$ is enforced to improve the UV behaviour of the theory\footnote{The same idea also works to construct (Lorentz violating but) renormalizable gauge theories in higher dimensions, see \textit{e.g.} \cite{Iengo:2009ix,Iengo:2010xg,Kanazawa:2014fla,Lambert:2022ztz}.}.
In these scenarios LI is fundamentally absent in the UV. In turn, a mechanism to enforce the emergence of LI in the IR is crucial for the phenomenological viability given the very stringent observational bounds on Lorentz violation \cite{Kostelecky:2008ts}. Several mechanisms are known to lead to accidental Lorentz symmetry at low energies.
A simple option is a non-relativistic form of supersymmetry \cite{GrootNibbelink:2004za}, see also \cite{Pujolas:2011sk,Redigolo:2011bv}. A more minimal option is to resort to the RG evolution itself, which turns out to lead to true emergence of LI in the IR, as noticed first in \cite{chadha1983lorentz,nielsen1978beta}, in the sense that the lightcone speed parameters of different species are driven to a common value. In weakly coupled theories, however, this RG flow is logarithmic \cite{Iengo:2009ix,Giudice:2010zb,Anber:2011xf}, in practice too slow for phenomenological viability.

This motivates the study of flows exhibiting the emergence of LI in strongly coupled theories\footnote{It is tempting to view graphene as a real-world example where LI emerges due to strong coupling, which arises close to the Dirac cone due to the inefficient screening and the small Fermi surface. Indeed, the fine structure constant in suspended graphene is $\alpha \approx 2.2$ (\textit{i.e.} not $\ll 1$). The focus of our work is to consider these flows in general, with no direct application to graphene.}. 
By exploiting holographic methods \cite{Natsuume:2014sfa,Baggioli:2019rrs}, Refs.\cite{Bednik:2013nxa,Kharuk:2015wga} have shown that strong coupling can indeed accelerate the RG flows towards a LI IR fixed point. In this work we expand the analysis aiming to assess the capabilities of this mechanism in more generality.

In order to be able to use holographic methods, our analysis will refer (as much as possible) to generic Conformal Field Theory (CFT). To trigger a LI breaking RG flow, we need at least one operator with integer spin. The simplest examples are vector operators. Conserved currents are `rigid', in that the scaling dimension $\Delta$ is fixed ($\Delta=3$ in $3+1$ dimensions). More interesting possibilities arise with non-conserved currents. For the sake of illustration, we will assume  one conserved and one non-conserved operators,
\bea
J^{(A)}_\mu && \quad \text{with} \quad \Delta=3\,,\nonumber \\
J^{(B)}_\mu && \quad \text{with} \quad \Delta>3~.
\eea
The flow triggered by a finite density of particles is the RG flow induced by deforming the Lagrangian with an operator
$\propto J^{A}_t$. This operator is relevant so it induces a flow from $z_{UV}=1$ to a different theory in the IR. At zero temperature, and in the minimal theory (dual to Einstein-Maxwell in AdS), one gets $z_{IR}\to\infty$.

More interestingly, we will explore deformations induced by $J^{B}_t$, how the induced flows depend on the scaling dimension and, especially, on the nonlinear structure of the theory. For $\Delta>4$, $J^{B}_t$ is irrelevant so there is a chance to have RG flows with $z_{IR}=1$ but which violate Lorentz invariance in the UV. These flows, then, display LI as an accidental/emergent symmetry in the IR. Moreover, it is certainly possible that the emergence of LI is more `efficient' than logarithmic. The efficiency of emergence is precisely parameterized by  the {\em irrelevance}, $\Delta-4$, of $J^{B}_t$~\footnote{From the IR perspective, the leading Lorentz violating effects are expected to be dominated by the so-called `LILVO' -- the least irrelevant Lorentz violating operator. In our models, $J^B_t$ will play the role of the LILVO.}. In order to reach emergent LI efficiently in the IR, large values (of order $1-2$) are needed for $\Delta-4$. In the models studied before \cite{Bray:1994zz,Bednik:2013nxa}, it turns out that the irrelevance $\Delta-4$ is rather limited, below $\approx 0.35$. We will see that turning on interactions allows to reach larger $O(1-2)$ values. 

Away from the emergence of LI, one realizes that the case $\Delta=4$, is of special interest too. In this case $J^B_\mu$ is (non-conserved but) marginal. As such, it can trigger interesting flows, for instance with a logarithmically running coupling constant. A more dramatic possibility worth examining is whether it is possible to arrange that the operator is exactly marginal. We will see that  it is certainly possible to design (or tune) the interactions in gravity dual to implement holographically an exactly marginal current $J^{B}_\mu$. Deformations of the CFT by $J^{B}_t$, then, lead to a {\em Lifshitz critical line}: a continuous line of fixed points with different dynamical exponent $z$ at each critical point.

We will model these CFTs using the standard gauge-gravity duality dictionary. 
The RG flow structure is geometrized in the radial dynamics along an extra-dimensional holographic coordinate which is ``dual'' to the RG energy scale \cite{Balasubramanian:1999jd}. For simplicity, we will only consider solutions breaking Lorentz boosts but preserving rotational symmetry. Holographic RG flows with broken LI and holographic fixed points with Lifshitz scaling have been constructed and considered in several instances in the past, \textit{e.g.}, \cite{Kachru:2008yh,Taylor:2008tg,Braviner:2011kz,Charmousis:2010zz,Giataganas:2017koz,Liu:2012wf,Korovin:2013bua,Singh:2013iba,Korovin:2013nha,Kachru:2013voa,Dey:2014yra,Burda:2014jca,Cremonini:2014pca,Taylor:2015glc}.
The holographic duals of the conserved/non-conserved currents $J^{(A)}_\mu$, $J^{(B)}_\mu$ are  massless/massive vector fields, respectively denoted as $A_\mu$, $B_\mu$. Similar models have already been employed in Refs. \cite{Braviner:2011kz,Bednik:2013nxa}. Here, we will generalize the model in several directions: (I) by considering a more general potential for the massive gauge field $B_\mu$, (II) by introducing finite temperature effects, (III) by allowing for finite charge density ($A_t\neq 0$) and (IV) by constructing holographic RG flows with intermediate scaling regimes.

Finally, there has been great interest in understanding LV RG flows in order to establish the possible existence of a monotonic function ``counting'' the number of degrees of freedom in absence of LI. This line of investigation has been strongly inspired by the famous $c$-function proposed by Zamolodchikov \cite{1986JETPL..43..730Z}, the Cardy's $a$-function \cite{CARDY1988749} and various holographic and quantum information generalizations \cite{Komargodski:2011vj,Freedman:1999gp,Myers:2010xs,Myers:2010tj,Casini:2006es,Ghasemi:2019xrl} which are valid for LI RG flows. It is far less clear if any analogous monotonic function can be defined in cases when the Lorentz group is broken. There have been several proposals in the literature \cite{Chu:2019uoh,Caceres:2022hei} with interesting applications regarding the identification of quantum critical points using these probes \cite{Baggioli:2020cld,Baggioli:2023ynu}. One particularly simple proposal is given in \cite{Caceres:2022hei} where a specific $a$-function is built and its monotonicity argued using a radial null energy condition (see also \cite{PhysRevD.106.046005,Liu:2012wf} for relevant previous results). Such a proposal has been tested using holographic p-wave superfluid models, where rotations are spontaneously broken, but to the best of our knowledge was never confronted with LV RG flow with broken boost symmetry. Here, we will confirm that the proposal of \cite{Caceres:2022hei} is valid also in more general situations and for a large class of holographic RG flows at finite temperature and charge density.\\

The paper is organized as follows. We introduce the holographic model in Section \ref{sec:model} and classify the fixed points with or without relativistic scaling symmetry in Sec.\ref{sec:fixed points}. In Sec.\ref{sec:RG flows}, we identify the relevant and irrelevant deformations around the fixed points and construct numerically the RG flows using holographic methods. Then, in Sec.\ref{sec:emergence}, we study the emergence of LI in these holographic flows. We discuss the existence of a critical line of Lifshitz fixed points in Sec.\ref{sec:superpotential}. Finally, in Sec.\ref{sec:a-function}, we prove the existence of an $a$-function which always monotonically decreases along the RG flows. We summarize our results in Sec.\ref{sec:conclusion} and collect the computational details in appendices \ref{app:A}-\ref{app:B}.

\section{Holographic RG Lifshitz flows}
\subsection{Model}
\label{sec:model}
We consider a generalized holographic Einstein-Maxwell-Proca model in $n+2$ dimensions described by the following action
\begin{equation}
\mathcal{S}\,=\,\int d^{n+2}x \,\sqrt{-g}\,\left[\,\mathcal{R}\,-\,\mathcal{F}^2\,-\,\mathcal{F}_B^2-V\left(B_\mu\,B^\mu\right)\,\right]\,,
\end{equation}
where $\mathcal{F}=dA$ and $\mathcal{F}_B=dB$ are the field strengths of the $U(1)$ Maxwell field $A_\mu$ and of the Proca vector $B_\mu$ respectively. $V$ is a generic scalar potential for the scalar form $B^2\equiv B^\mu B_\mu$ which contains a constant and negative cosmological constant, $V(0)=-n(n+1) <0$.

In the spirit of Ref. \cite{Megias:2014iwa}, we write down the ansatz for the metric and matter fields in a convenient ``RG-Gauge'' given by
\bea
\begin{split}
\label{Ansatz}
& ds^2\,=\,-e^{2u}\,dt^2+ e^{2\int_0^u w(\xi)\,d\xi} \sum_{i=1}^n dx_i^2\,+\,N(u)^2 \,du^2\,\,,\\
& A\,=\,A_t\,dt\,=\,\left(\int_0^u \alpha(\xi)\,e^\xi\,N(\xi)\,d\xi\right)\,dt\,,\\
& B\,=B_t\,dt\,=\,\,e^u\,\beta(u)\,\,dt\,,
\end{split}
\eea
where the radial coordinate $u$ spans from the UV boundary $u_{\text{UV}}\equiv+\infty$ to the IR horizon $u_h\equiv -\infty$ and $\{x_1, x_2, ...x_n\}$ defines an $n$-dimensional space with $\mathrm{SO(n)}$ rotational symmetry. Note that in these variables, $w(\xi)$ plays the role of the inverse dynamical exponent. On fixed points, $w(\xi)$ is constant and one identifies the  Lifshitz exponent (see Eq. \eqref{ll}) as
\begin{equation}
z \equiv 1/w_0\;.
\end{equation}

A finite value for any of $A_t$ or $B_t$ implies the breaking of Lorentz invariance. Similarly, one can switch on spatial components for these vectors, $A_i$ or $B_i$, to break rotational symmetry. We will not consider this scenario here, see for example \cite{Cremonini:2014pca,Baggioli:2023yvc}. Also, we will keep translational invariance unbroken.

The corresponding equations of motion for a general potential $V$ take a complicated form, and can be found in Appendix \ref{app:A}. The temperature of the dual field theory can be defined through the surface gravity at the horizon as
\begin{equation}
T_H\,=\,\frac{\kappa}{2\pi}\,=\,\frac{1}{2\pi}\sqrt{-\frac{n^{\mu;\nu}n_{\mu;\nu}}{2}}\,,
\end{equation}
with $n^\mu$ a timelike killing vector.
For our ansatz in Eq.\eqref{Ansatz}, the temperature is explicitly given by
\begin{equation}
\label{TEMP}
T_H\,=\,\left.\frac{e^u}{2\,\pi\,N(u)} \right\vert_{u_h}\,,
\end{equation}
where the location of the horizon $u_h$ is defined from $g_{tt}=e^{2u}=0$. At zero temperature, $T=0$, the horizon locates at $u_h\rightarrow -\infty$ with $N(u_h)$ a non-zero constant. At finite temperature, $T>0$, the horizon locates at $u_h\rightarrow -\infty$, and $N(u)\rightarrow 0$ as $N(u)\sim e^u$ when $u\rightarrow -\infty$.

\subsection{Fixed point taxonomy}
\label{sec:fixed points}

The model described in the previous Section admits a large class of RG flow solutions interpolating between various Lorentz invariant and Lorentz violating fixed points with either zero or finite charge and temperature. By fine-tuning the boundary conditions, non-trivial intermediate scaling regimes can also be achieved. 

We start by classifying all the possible fixed points of the system admitted by the bulk equations of motion, Eqs.\eqref{GenEQs} in Appendix \ref{app:A}. By definition, fixed points are described by configurations for which all the ansatz functions take constant values $\{\alpha_0,N_0,w_0,\beta_0\}$. According to the ansatz in Eq.\eqref{Ansatz}, we obtain the equations of motion for the fixed points without specifying any concrete form for the potential $V$. Since they are rather complicated and their form is not particularly illuminating, we present them in Eqs.\eqref{FixP} in Appendix \ref{app:A}.

There are four types of possible fixed points admitted by our model that are solutions of Eqs.\eqref{FixP}. They do correspond to four different types of geometries:
\begin{enumerate}
\item\label{uno} AdS$_{n+2}$ spacetime;
\item\label{due} 
Near-horizon geometry, dual to deforming the theory by introducing finite temperature. 
\item Charged near-horizon geometry, denoted in the rest of the manuscript as ``Q-horizon'', and dual to a deformation by finite temperature and finite density \footnote{There is no real scaling symmetry (fixed point) in the IR in this nor the previous case, since these solutions have finite temperature.}.  
\item\label{tre} Extremal charged near-horizon, AdS$_{2}\,\times\,R^n$, geometry (corresponding to finite density, $T=0$ deformation);
\item\label{quattro} Lifshitz spacetime. 
\end{enumerate}
Since we perform Lorentz breaking deformations, it is generically possible to end on an ``IR fixed point" that corresponds to a black brane solution.

For simplicity, we first focus on the subclass of fixed points with vanishing Proca Field, $\beta(u)=0$ and a constant potential $V(z)=V_0$ which defines the (negative) \textit{cosmological constant}. This class includes the geometries labelled with indices 1,2,3 above. The equations of motion in these cases boil down to the following algebraic equations,
\begin{align}
&N_0^2=\frac{n w_0\left[(n-1) w_0+2\right]}{V_0-2 \alpha_0^2}\,,\nonumber\\
&0=\frac{n w_0 ((n-1) w_0+2) \left(V_0(w_0-1) ((n-1) w_0+1)+2 \alpha_0^2 \left(w_0 \left((n-1)^2 w_0+3
   n-2\right)+1\right)\right)^2}{V_0-2 \alpha_0^2}\,,\nonumber\\
&0=\frac{w_0 ((n-1) w_0+2) \left(2 \alpha_0^2 ((n-1) w_0+1)+V_0 (w_0-1)\right)}{V_0-2 \alpha_0^2}\,,\nonumber\\
&0=n\,\alpha_0 w_0\,.
\end{align}

The first and obvious fixed point of the system is Anti-de Sitter spacetime, which corresponds to the following choice of parameters:
\begin{equation}
\text{AdS}_{n+2}\, := \,\{\alpha_0\,=\,0,\, \beta_0\,=\,0,\, w_0\,=\,1,\,N_0\,=\,1\}\,,
\end{equation}
where both bulk gauge fields vanish. The AdS$_{n+2}$ geometry is supported by a constant valued potential that is identified with the negative cosmological constant,
\begin{equation}
V_0=-n(n+1)\,.
\end{equation}
It is simple to verify that $\text{AdS}_{n+2}$ has no temperature, \textit{i.e.}, $T=0$. Also, the fixed point in the dual field theory enjoys the full Lorentz group and indeed the whole conformal group in $n+1$ dimensions.

The second ``fixed point'' of the system is given by:
\begin{equation}
\{\alpha_0\,=\,q,\, \beta_0\,=\,0, \,w_0\,=\,0,\,N_0\,=\,0\}\,.
\end{equation}
This case, labelled with index 2 above, corresponds to the near-horizon geometry of a black hole at finite temperature, where the free parameter $q$ stands for the electric charge of the corresponding black hole. One can show (for $q=0$ even analytically) that as $u\rightarrow -\infty$, $N(u)$ together with $w(u)$ approach to zero exponentially, \textit{i.e.} $N(u),w(u) \sim e^u$, such that the temperature in Eq.\eqref{TEMP} is finite. To distinguish the charge $q \neq 0$ and the neutral cases, in the rest of the manuscript, we will refer to the near-horizon geometry with $q \neq 0$ as ``Q-horizon'' and to the $q=0$ one as ``near-horizon''. As noted above, there is no real scaling symmetry (hence, no real fixed points) in this case, apart from the extremal case, \textit{i.e.}, the $T=0$ limit of the Q-horizon geometry.
This geometry corresponds to a dual field theory at finite temperature and finite charge density.

Finally, it is possible to obtain a third geometry with $w_0=0$ but $N_0$ finite. This happens when the electric charge reaches a special ``extremal`` value satisfying
\begin{equation}
-n\,(n+1)\,+\,2\,\alpha_0^2\,=\,0\,.
\end{equation}
This expression represents an extremality condition and the corresponding solution is  the AdS$_2\,\times R^n$ near-horizon geometry of a charged black hole at zero temperature. In this case, the fixed point is represented by the following parameters
\begin{equation}
\text{AdS}_2\,\times \text{R}^n\,:=\,\left\{\alpha_0\,=\,\sqrt{\frac{n\,(n+1)}{2}},\, \beta_0\,=\,0, \,w_0\,=\,0,\,N_0\,=\,\frac{1}{\sqrt{(n+1)\,n}}\right\}\,,
\end{equation}
where the value of $N_0$ is related to the radius of the AdS$_2$ geometry. This third geometry corresponds to a field theory at zero temperature but finite charge density.
  
We now turn to the study of the fourth class of fixed points which exhibit Lifshitz scaling \eqref{ll} and which are supported by a non zero value $\beta_0$ for the gauge field $B_\mu$ along with a non trivial potential $V\left(B_\mu B^\mu\right)$. In the bulk description, this class corresponds to the Lifshitz spacetime labelled with index 4 above.
Setting the electric charge $\alpha_0$ to zero, we are left with the following algebraic system:
\bea
\begin{split}
0&=2 \,\beta_0^2+\frac{n \,(n+2\, z-1)}{z^2}+N_0^2\, V\left(-\beta_0^2\right)\,,\\
0&=\frac{n \left[\,(z-1)\, (n+z-1)-2 \,\beta_0^2 \,z^2\,\right]}{\beta_0^2 \,(n-1) \left[\,n\, (n+2 \,z-1)+2 \,\beta_0^2\,
   z^2\,\right]}+\frac{V'\left(-\beta_0^2\right)}{V\left(-\beta_0^2\right)}\,,\\
   0&=\frac{2 \,z^2}{n \,(n+2\, z-1)+2\, \beta_0^2\, z^2}+\frac{1-z}{\beta_0^2 \,(n+z-1)}-\frac{V'\left(-\beta_0^2\right)}{V\left(-\beta_0^2\right)}\,,\\
0&=2 \left(\frac{1}{2 \,\beta_0^2\, (n-1)-n}+\frac{z^2}{n\, (n+2 \,z-1)+2\, \beta_0^2\, z^2}\right)-\frac{V'\left(-\beta_0^2\right)}{V\left(-\beta_0^2\right)}\,,
\label{Lif}
\end{split}
\eea
where we identify $w_0=1/z$ with the Lifshitz scaling parameter $z$ and define $V'\equiv \frac{\partial V}{\partial(-\beta_0^2)}$. The first equation in Eqs.\eqref{Lif} is a constraint which uniquely defines $N_0$ in terms of the other free parameters. We first concentrate on the other equations.
Summing up the second and third equations in Eqs.\eqref{Lif}, we obtain the simple expression
\begin{equation}
\frac{2}{2 \,\beta_0^2 \,(n-1)-n}+\frac{z-1}{\beta_0^2\, (n+z-1)}\,=\,0\,
\end{equation}
that leads to the relation between the Proca charge $\beta_0$ and the Lifshitz exponent $z$
\begin{equation}
\beta_0\,=\,\sqrt{\frac{z-1}{2\,z}}\,.\label{betalif}
\end{equation}
Remarkably, this expression is independent of the form of the potential $V(B_\mu B^\mu)$. Further, as shown in Eq.\eqref{betalif}$, \beta_0\geq 0$ is consistently equivalent to $z\geq 1$ \cite{Hoyos:2010at}. Using the solution \eqref{betalif}, there is only one more independent equation that relates the form of the potential, and all the possible free parameters entering therein, to the Lifshitz dynamical exponent $z$,
\begin{equation}
\frac{2\, n\, z}{(n+z-1) \,(n+z)}+\frac{V'\left(\frac{1}{2} \left(\frac{1}{z}-1\right)\right)}{V\left(\frac{1}{2} \left(\frac{1}{z}-1\right)\right)}\,=\,0\,.
\end{equation}
As a warm-up, we first review the case of the linear potential
\begin{equation}
V(\zeta)\,=\,V_0\,+2 \,m^2\,\zeta\,,
\label{LinPot}
\end{equation}
where $\zeta\equiv B_\mu B^\mu = -\beta_0^2$, which was previously analyzed in \cite{Braviner:2011kz}. Within this simple choice, the Lifshitz fixed points are defined by
\begin{align}
N_0\,=\,\sqrt{\frac{n^2+n \,z+z^2-z}{ z^2\,(n^2+n)}}\,,\quad m^2\,=\,\frac{n^2 (n+1) z}{n^2+n \,z+(z-1) z}\,.
\label{LifLin}
\end{align}
\begin{figure}[t]
\centering
\includegraphics[width=65mm]{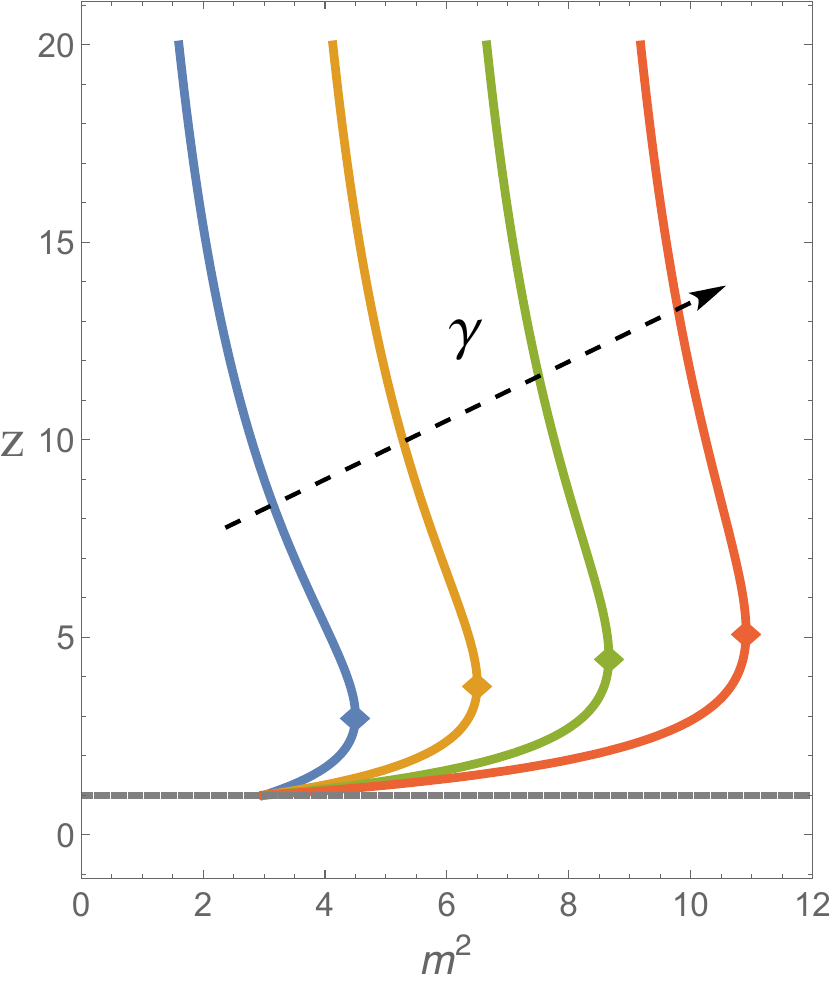}
\caption{The two branches of Lifshitz fixed points for $n=3$ and the quadratic potential \eqref{quadrpot} increasing the parameter $\gamma=0,5,10,15$. Note that the curve for $\gamma=0$ corresponds to the linear potential case considered in \cite{Braviner:2011kz}. Here, $m$ is the mass of the Proca bulk field $B_\mu$. The diamond symbols characterize the separation between the two branches of Lifshitz fixed points and correspond to the maximum dynamical exponent achievable in the UV $z_{max}$ (see more below).}
\label{lifFix}
\end{figure}
More in general, we can think of the potential as a polynomial in $\zeta$ given by
\begin{align}
V(\zeta)\,=\,\underbrace{V_0\,+2 \,m^2\,\zeta}_{\text{Ref.\cite{Braviner:2011kz}}}\,+\,\sum_{k=2}^{N} \,\tilde{c}_k\,\zeta^k\,=\,\sum_{k=0}^{N} \,c_k\,\zeta^k.
\label{PGen}
\end{align}
For simplicity, in the paper, we will mostly focus on the simplest generalization which consists in truncating the series of Eq.\eqref{PGen} up to the second order. 
We will therefore consider a potential of the form
\begin{equation}
V(\zeta)\,=\,-n(n+1)+2\,m^2\,\zeta+\gamma\,\zeta^2\,,
\label{quadrpot}
\end{equation}
which can be thought as a quartic potential for the Proca vector $B_\mu$.
The fixed points with Lifshitz scaling can be found to satisfy
\bea
\begin{split}
\label{Lif2nd}
&m^2\,=\,\frac{4 \,n^2 (n+1) z^2+\gamma  (z-1) \left(2\, n^2+n\, (3\, z-1)+2 \,(z-1) \,z\right)}{4 \,z \left(n^2+n \,z+(z-1)\, z\right)}\,,\label{mquadr}\\
&N_0\,=\,2 \,\sqrt{\frac{n^2+n\, z+(z-1)\, z}{4\, n\, (n+1)\, z^2+\gamma\,  (z-1)^2}}\,,
\end{split}
\eea
which reduce to the solutions in Eq.\eqref{LifLin} in the limit $\gamma=0$. Notice that for $\gamma>0$ the mass is strictly real ($m^2>0$) and there exist no instabilities. In Fig.\ref{lifFix}, we illustrate the solutions for several values of $\gamma$ and make a comparison with the linear potential case already considered in \cite{Braviner:2011kz}. One can clearly find that $z_{max}$, defined as the maximum $z$ allowed (and indicated with colored diamond symbols in Fig.\ref{lifFix}), increases monotonically as a function of $\gamma$.

We summarize the different types of fixed points present in our setup in Table \ref{tfix}, where the corresponding values of the model parameters are indicated.

\begin{table}[h!] 
  \centering
  \begin{tabular}{c|cccc}
    \toprule
    %$\alpha=0$   & 
     Fixed Points & $\beta_0$ & $\alpha_0$ & $w_0$ & $N_0$\\
    \midrule
    %\checkmark &   $\leq q^2/3$ & \checkmark & 0 at $\theta=q^2/3$ & $\infty$ & 0\\
    %\checkmark & $> q^2/3$ & $\backslash$ & $\backslash$ & $\backslash$ & $\backslash$\\
    %$\backslash$ & 
    AdS & 0  & 0 & 1 & 1\\[2mm]
    %$\backslash$ & 
    near-horizon & 0 & 0 & 0  & 0\\[2mm]
    Q-horizon & 0 & $q$ & 0  & 0\\[2mm]
    %$\backslash$ & 
    $\text{AdS}_2\,\times \text{R}^n$ & 0 & $\sqrt{\frac{n\,(n+1)}{2}}$ & 0 & $\frac{1}{\sqrt{(n+1)\,n}}$
\\[2mm] 
Lifshitz & $\sqrt{\frac{z-1}{2\,z}}$ & 0 & $1/z$ & $\star_V$
\\[2mm]    \bottomrule
  \end{tabular}
  \vspace{0.3cm}
      \caption{The fixed points present in our model. We show the values of the various functions at the fixed points. As analyzed in more detail below, $\star_V$ means that this value depends on the choice of the potential $V\left(B_\mu B^\mu\right)$. The parameter $z$ is the Lifshitz dynamical exponent in Eq.\eqref{ll}.
  }
  \label{tfix}
  \end{table}
\subsection{Linear stability}
\label{sec:RG flows}
In the previous Section, we have discussed all the possible fixed points in our system including the conditions for their existence and their scaling symmetries. The second step of our analysis is to determine the linear stability of the various fixed points and the nature of possible relevant deformations. This will allow us to predict and construct RG flows between different fixed points. From the gravitational viewpoint, a RG flow in the dual field theory corresponds to a bulk geometry with different UV and IR asymptotic behaviors. In common jargon, the RG flow is geometrized and emerges from the dynamics along the holographic extra dimension, which plays the role of the RG scale.

The analysis in this Section is focused on the theory with potential
\begin{equation}
V(\zeta)\,=\,-n(n+1)+2\,m^2\,\zeta+\gamma\,\zeta^2\,.
\end{equation}
Since the equations of $N(u)$ and $N'(u)$ are completely determined once the profiles for the other fields are known, we only consider the perturbations for $w(u), \alpha(u)$ and $\beta(u)$ using the following ansatz
\bea
w(u)=w_0+\delta w(u)\,,\quad 
\alpha(u)=\alpha_0+\delta \alpha(u)\,,\quad
\beta(u)=\beta_0+\delta \beta(u)\,.
\eea
Our analysis will be limited to the linear dynamics, at leading order in the above perturbations. This will be enough to reveal the stability of the different fixed points for small deformations around them.\\

\noindent {\bf AdS$_{n+2}$.} We start by considering the case of the AdS$_{n+2}$ fixed point defined by the following leading order solution 
\bea
\{w_0, \alpha_0, \beta_0\}=\{1,0,0\}\,.
\eea
We find that the linear perturbations around AdS$_{n+2}$ obey the following behaviors
\begin{align}
   & \delta \alpha (u)\,=\,c_1\,e^{-n\,u},\quad \delta w (u) \,=\,c_2 \,e^{-(n+1)\,u},\quad \delta \beta (u)\,=\,c_3 \,e^{-u\,\Delta_{\beta}^{+}}\,+\,c_4\,e^{-u\,\Delta_{\beta}^{-}} 
\end{align}
with:
\begin{align}
\Delta_{\beta}^{\pm}\,=\,\frac{1}{2}\,\left(n+1\pm\sqrt{2\, V'(0)+(n-1)^2}\right)\,=\,\frac{1}{2} \left(n+1\pm\sqrt{4 \,m^2+(n-1)^2}\right).
\nonumber\\
   & \label{procadim}
\end{align}
It is simple to verify that $n,n+1,\Delta_\beta^+$ are always positive, while $\Delta_\beta^->0$ when $m^2<n$, $\Delta_\beta^-<0$ when $m^2>n$. The negative powers in the exponential functions represent unstable directions towards $u\rightarrow -\infty$ (IR), while on the contrary the positive powers correspond to growing perturbations towards the UV, $u\rightarrow +\infty$. In $B_t=e^u\beta(u)$ there is an additional factor $e^u$, however, the perturbation of the Proca field contributes to the Lagrangian as $\beta^2$ and $(\beta+\beta')^2$, and thus the convergence or divergence nature of this perturbation can be read out from $\beta(u)$ instead of $B_t(u)$. 
All in all, we conclude that,
\begin{itemize}
\item when $m^2<n$, the perturbations of the Proca field are always relevant and the system can only flow away from the AdS fixed point. This implies that the AdS fixed point can only be a UV fixed point since it is not IR attractive.
\item When $m^2>n$, there are both relevant and irrelevant directions around the AdS fixed point. Therefore, in this range, the AdS fixed point can be both an IR and a UV fixed point.
\end{itemize}

\noindent {\bf Lifshitz.} For the sake of clarity, let us consider the simpler case $\gamma=0$. The leading order solution defining the fixed point is 
\bea
\{w_0,\alpha_0,\beta_0\}=\left\{\frac{1}{z},0,\sqrt{\frac{z-1}{2z}}\right\},
\eea
and the linear perturbations around it follow 
\bea
\begin{split}
\delta\alpha(u)&=\alpha_1~e^{-\frac{n}{z}u}\,,\\
\delta w(u)&=w_1~e^{-(1+\frac{n}{z})u}
+w_2~e^{-\frac{\Delta^+_\beta}{z}u}+w_3~e^{-\frac{\Delta^-_\beta}{z}u}\,\\
\delta \beta(u)&=\beta_1~e^{-(1+\frac{n}{z})u}
+\beta_2~e^{-\frac{\Delta^+_\beta}{z}u}+\beta_3~e^{-\frac{\Delta^-_\beta}{z}u}\,,
\end{split}
\eea
where
\bea
\Delta^{\pm}_\beta=\frac{1}{2}\left(n+z\pm\sqrt{9z^2-(6n+8)z+n(n+8)}\right)\,.\label{lsls}
\eea
Similar to the AdS$_{n+2}$ case, $n/z, 1+n/z$ and $\Delta^+_\beta$ are always positive, while $\Delta^-_\beta>0$ when $z<n$, $\Delta^-_\beta<0$ when $z>n$. The negative powers in the exponential functions decay at $u\rightarrow +\infty$, while the positive at $u\rightarrow -\infty$. Then, we conclude that,
\begin{itemize}
\item when $z<n$, the perturbations are always relevant and the RG trajectories can only flow away from the Lifshitz fixed point. In this range, the Lifshitz fixed point can only be a UV fixed point.
\item When $z>n$, there are both relevant and irrelevant directions around the Lifshitz fixed point. Here, both IR and UV Lifshitz fixed point are admitted.
\end{itemize}
The same analysis as above can be performed analytically for $\zeta \neq 0$ providing a straightforward generalization of Eq.\eqref{lsls}. Nevertheless, the expressions are rather lengthy and not so useful and therefore not shown explicitly here.\\

\noindent{\bf Q-horizon.} The leading order solution defining the Q-horizon fixed point is given by
\bea
\{w_0,\alpha_0,\beta_0\}=\{0,q,0\}\,,
\eea
and the linear perturbations around it by
\bea
\begin{split}
\delta\alpha(u)&=\frac{c_1}{2}\left(1-e^{2u}\right)n\,q+c_2\,,\\
\delta w(u)&=c_1~e^{2u}\,,\\
\delta \beta(u)&=c_3~e^u+c_4~e^{-u}\,.
\end{split}
\eea
All perturbations apart from the one parameterized by the coefficient $c_4$ are IR attractive and decay at $u\rightarrow -\infty$. Because of regularity at the IR horizon, one has to set $c_4=0$ to make sure that $B_t$ vanishes as $u \rightarrow-\infty$.\\

\noindent {\bf AdS$_2\times$ R$^n$.} The leading order solution is given by 
\bea
\{w_0,\alpha_0,\beta_0\}=\{0,\sqrt{\frac{n(n+1)}{2}},0\}\,,
\eea
and the linear perturbations around AdS$_2\times$ R$^n$ take the following form
\bea
\begin{split}
\delta\alpha(u)&=\frac{c_1}{2}\left(1-e^{2u}\right)n\sqrt{\frac{n(n+1)}{2}}+c_2\,,\\
\delta w(u)&=c_1~e^{2u}\,,\\
\delta \beta(u)&=c_3~e^u+c_4~e^{-u}\,.
\end{split}
\eea
It is straightforward to verify that the stability properties of this fixed point are identical to that of the Q-horizon case. This fixed point is therefore always IR attractive.\\

So far, we have determined the stability of the fixed points from a purely bulk gravitational perspective. Here, we briefly show how to recover the same results from pure field theory arguments. For completeness, in Appendix \ref{app:B}, we remind the reader about the bulk-boundary dictionary for holographic systems violating Lorentz invariance.\\

\noindent {\bf AdS$_{n+2}$.} Let us set $r\equiv e^u$ to make a direct comparison with the computations in Appendix \ref{app:B}, and write
\bea
\delta B_t(u)=c_4~r^{1-\Delta_\beta^-}+c_3~r^{1-\Delta_\beta^+}\,,
\eea
where the first leading term plays the role of source for the operator dual to the bulk field $B_t$ in standard quantization. Then, the mass dimension of the operator dual to the Proca field is 
\bea
1-\Delta_\beta^-+n=\Delta_\beta^+\,.
\eea
Therefore, we can immediately deduce, in agreement with the previous bulk analysis, that,
\begin{itemize}
\item $\Delta^+_\beta<n+1$ or $m^2<n$: the Proca field induces relevant directions. 
\item $\Delta^+_\beta=n+1$ or $m^2=n$: this term is marginal. 
\item $\Delta^+_\beta>n+1$ or $m^2>n$: the Proca field induces irrelevant directions. 
\end{itemize}

\noindent{\bf Lifshitz.} Following the same strategy, we write
\bea
\delta B_t(u)=\beta_3~r^{1-\frac{\Delta^-_\beta}{z}}+\dots
\eea
where we keep only the source term in the standard quantization. Then, the mass dimension of the operator dual to the Proca field is 
\bea
z-\Delta^-_\beta+n=\Delta^+_\beta\,.
\eea
We can find that
\begin{itemize}
\item $\Delta^+_\beta<n+z$ or $z<n$: the Proca field induces relevant directions.  
\item $\Delta^+_\beta<n+z$ or $z=n$: this term is marginal.
\item $\Delta^+_\beta>n+z$ or $z>n$: the Proca field induces irrelevant directions. 
\end{itemize}
This is again in perfect agreement with our previous results.\\

\noindent {\bf AdS$_2\times$ R$^n$.} In this case, a CFT$_1$ emerges as the dual field theory description of the AdS$_2$ geometry. Then, we need to analyze the stability in terms of a CFT$_1$. 
Using $r\equiv e^u$, we obtain 
\bea
\delta B_t(u)=c_4~r^2+c_3~r^0\,,
\eea
where the first term plays the role of source in standard quantization. Then, the mass dimension of the operator dual to the Proca field is $=2$. Since $\Delta_\beta=2>n_{\text{IR}}=0$, this term only induces irrelevant directions and AdS$_2\times$ R$^n$ can only be the IR fixed point.

In summary, our linear stability analysis tells us that the AdS$_{n+2}$ and Lifshitz$^-$ branches can serve as fixed points in the UV, while Lifshitz$^+$, AdS$_2\times$ R$^n$, near-horizon and AdS$_{n+2}$ can be IR fixed points. Finally, AdS$_{n+2}$ and Lifshitz$^+$ can appear also as intermediate scaling geometries. 

\begin{figure}[t]
\centering
\includegraphics[width=0.39\textwidth]{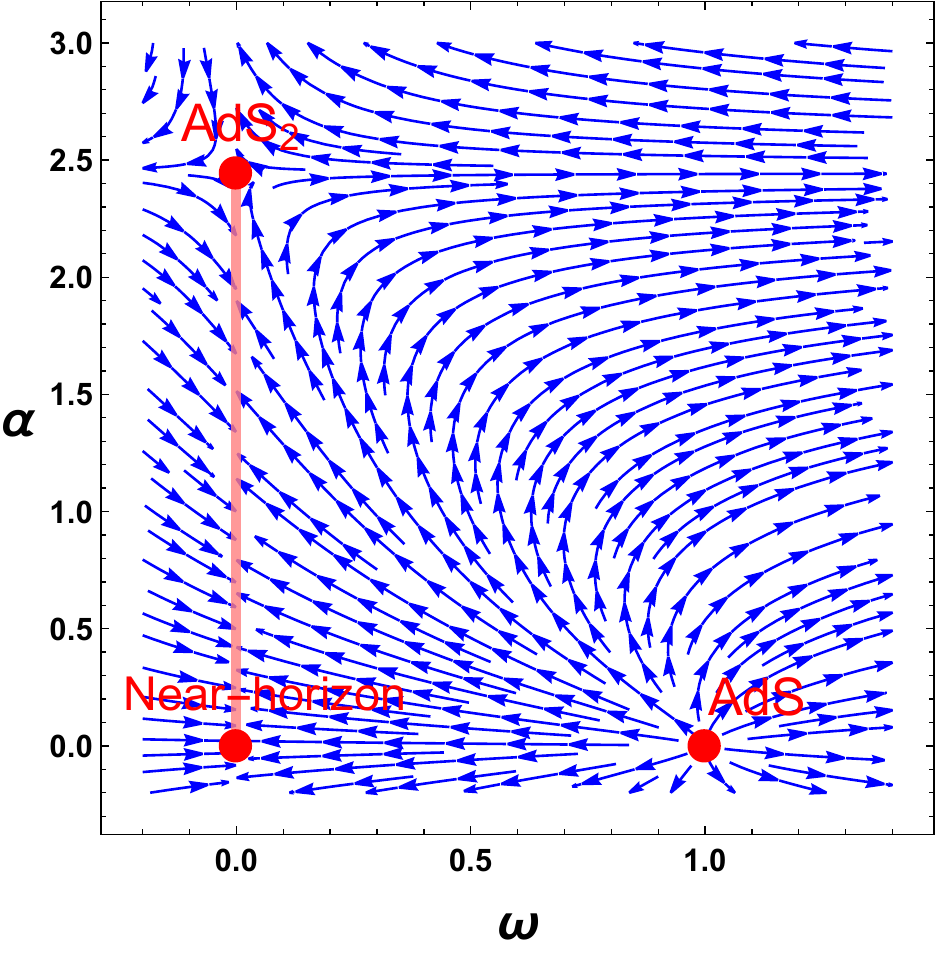}\hspace{0.5cm}
\includegraphics[width=0.4\textwidth]{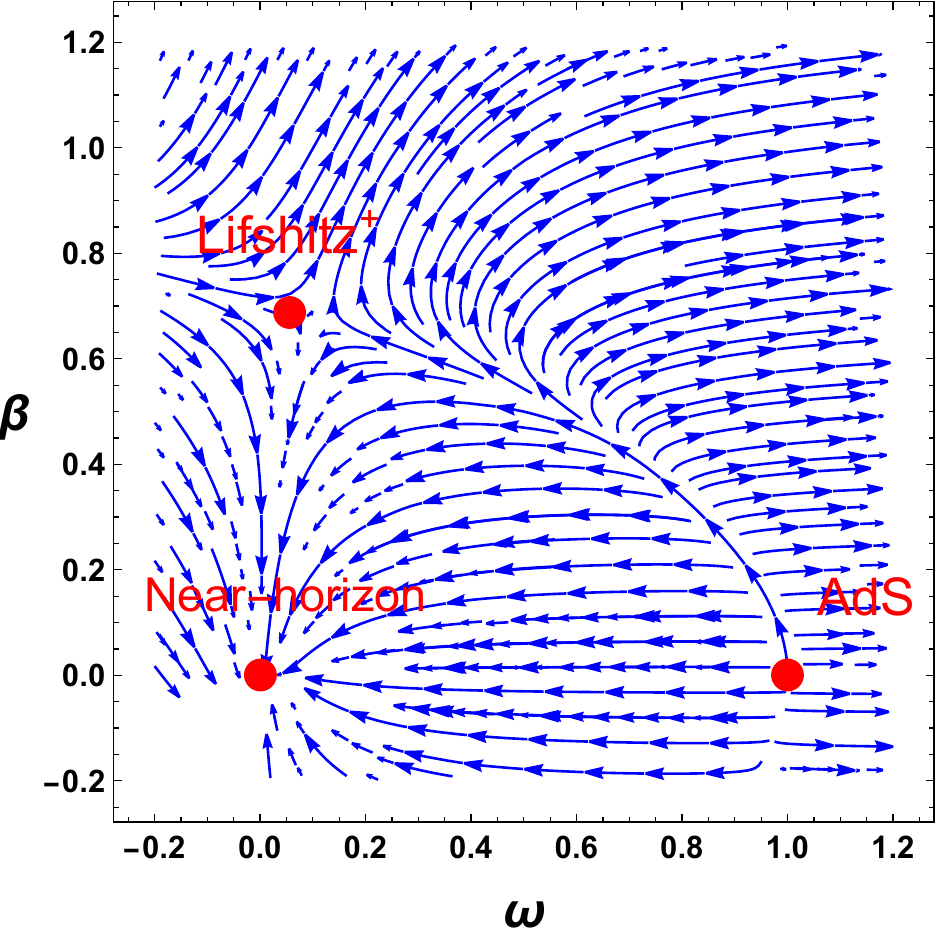}\\
\vspace{0.5cm}
\includegraphics[width=0.4\textwidth]{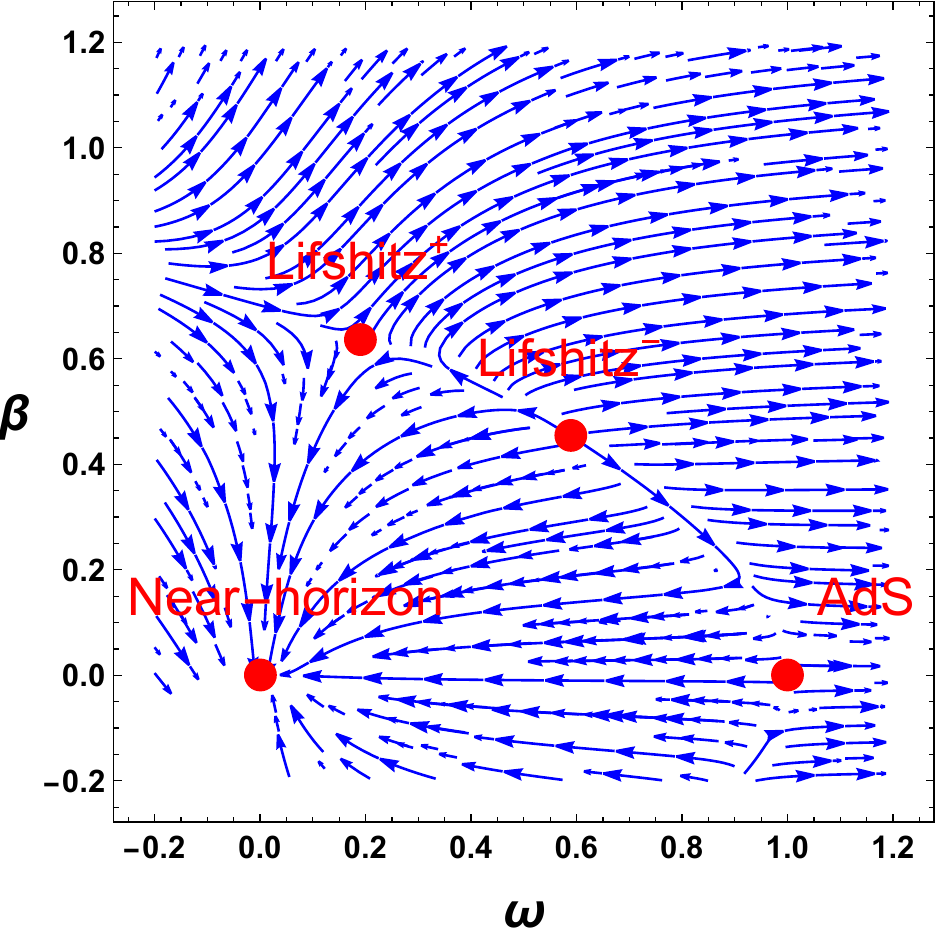}\hspace{0.5cm}
\includegraphics[width=0.4\textwidth]{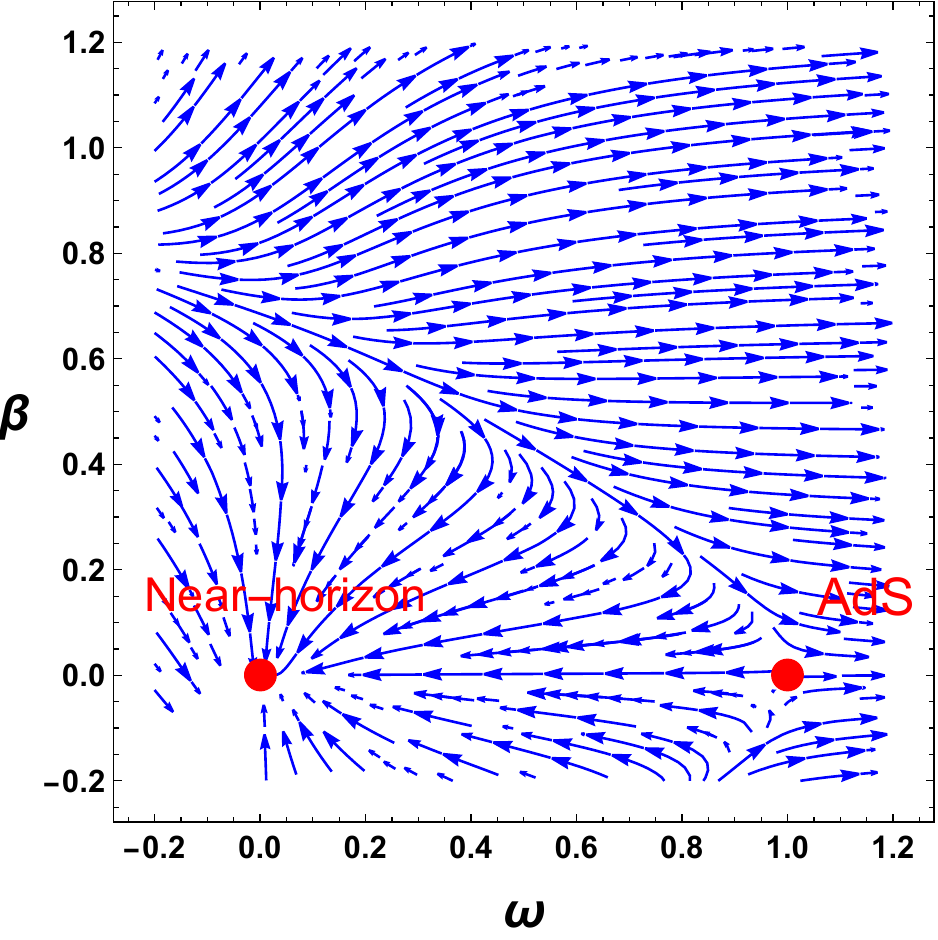}
\caption{RG flow phase diagram (blue arrows) around the fixed points (red points), with the in-going arrows indicating the irrelevant directions and out-going arrows the relevant ones. These flows are obtained from the linearized RG flow equations and thus only reliable around the fixed points indicated in red. 
{\bf Top left}: $\beta=0$. The fixed points in the $\alpha-\omega$ plane are AdS$_{n+2}$ and a continuous branch of Q-horizon solutions (pink thick line) starting from neutral near-horizon ($q=0$) to AdS$_2$ ($q=q_e$). 
{\bf Top right}: $\alpha=0, m^2=2.8$. The fixed points include near-horizon, AdS$_{n+2}$ and Lifshitz$^+$. 
{\bf Bottom left}: $\alpha=0, m^2=4$. The fixed points include near-horizon, AdS$_{n+2}$ and Lifshitz$^\pm$. 
{\bf Bottom right}: $\alpha=0, m^2=7$. The fixed points include only near-horizon and AdS$_{n+2}$. }
\label{fig:streams}
\end{figure}

To visualize this structure better, we consider the RG linearized trajectories in two-dimensional subspaces in Fig.\ref{fig:streams}. In these plots, to accommodate the plot in two dimensions we have either set the Proca parameter $\beta=0$ ({\em top left panel}) or the electric charge to zero ({\em other three panels}). The blue lines point along the directions of the RG flows in the $\beta=0$ or $\alpha=0$ planes. We emphasize that this linear analysis, and the corresponding vector plots in Fig.\ref{fig:streams}, are valid only around the fixed points. Away from them, non-linear terms, that are neglected in this approximation, are not negligible anymore and one has to resort to a full numerical analysis, as done in the next Sections. 

In the top left panel of Fig.\ref{fig:streams}, we show the situation with $\beta=0$ that does not allow for Lifshitz fixed points. In there, the continuous pink line between the AdS$_2$ and near-horizon geometries represents a continuous line of Q-horizon solutions. All the blue streams come out from the AdS$_{n+2}$ fixed point. This indicates that the AdS$_{n+2}$ can only be the fixed point in the UV once we set $\beta(u)=0$, since all the directions around it are relevant. On the contrary, the arrow can only flow into the points along the pink line, and this indicates that the near-horizon, Q-horizon and the AdS$_2\times$ R$^n$ fixed points are only allowed as IR fixed points, since all directions around them are attractive. 

In the other three panels in Fig.\ref{fig:streams}, we consider the situation with zero charge density, $\alpha_0$, but finite $\beta$ that allows for the presence of fixed points with Lifshitz scaling. We show respectively three examples with $m^2$ larger, equal and smaller than $n$. The differences between these three cases are already apparent in these vector plots and will be made even clearer below where the full RG flow solutions will be constructed numerically.

\subsection{Examples and classification of RG flows}
\label{sec:RG examples}
As analysed in the previous Section, the near-horizon, Q-horizon and the AdS$_2\times$ R$^n$ fixed points can only serve as IR fixed points, on the contrary the Lifshitz$^-$ can only appear as UV fixed point. Due to the existence of both relevant and irrelevant deformations near the AdS$_{n+2}$ and Lifshitz$^+$ fixed points, they can serve as UV fixed points but appear also as intermediate metastable fixed points. In this Section, we consider $n=3$ and numerically construct all possible RG flows, which are classified and summarized in Table \ref{table:list}. 

Before showing the examples below, let us introduce the classification method and the notations used for clarity. In each figure, the UV and intermediate geometries are fixed to be the same, while the RG flows indicated with different colors correspond to different IR fixed points. The color scheme is as follows: red is for thermal ($T\neq 0$) solutions, green is for charged solutions ($q\neq 0$), and yellow is for both thermal and charged. Moreover, blue lines represent the Lifshitz$^+$ IR fixed point with finite $\beta$, while black lines are for the vacuum state, \textit{i.e.}, the AdS$_{n+2}$ IR fixed point. In Table \ref{table:list}, we follow the same classification and notations. Finally, in the plots we indicate $u=\pm \infty$ respectively as UV and IR.

\begin{figure}[h]
\centering
\includegraphics[width=0.45\textwidth]{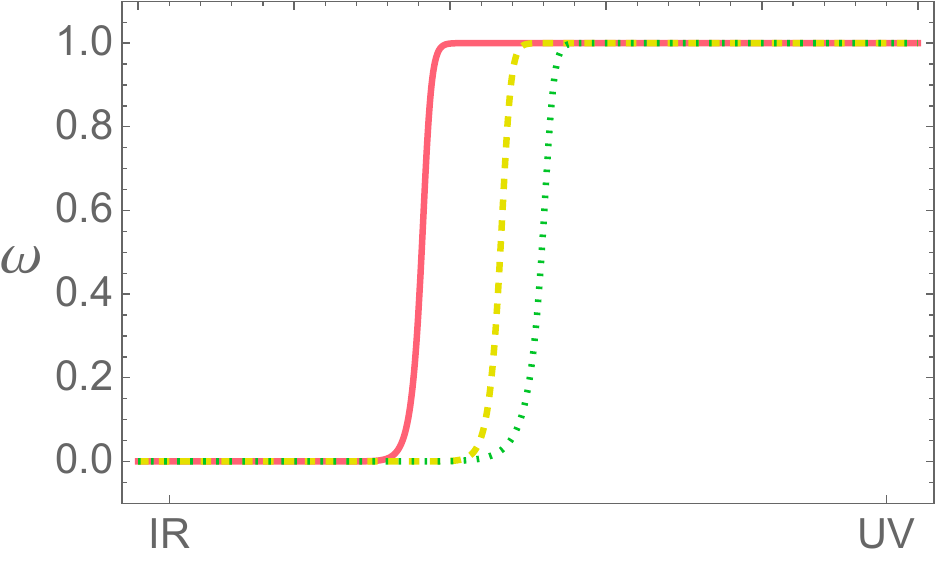}
\hspace{1cm}
\includegraphics[width=0.45\textwidth]{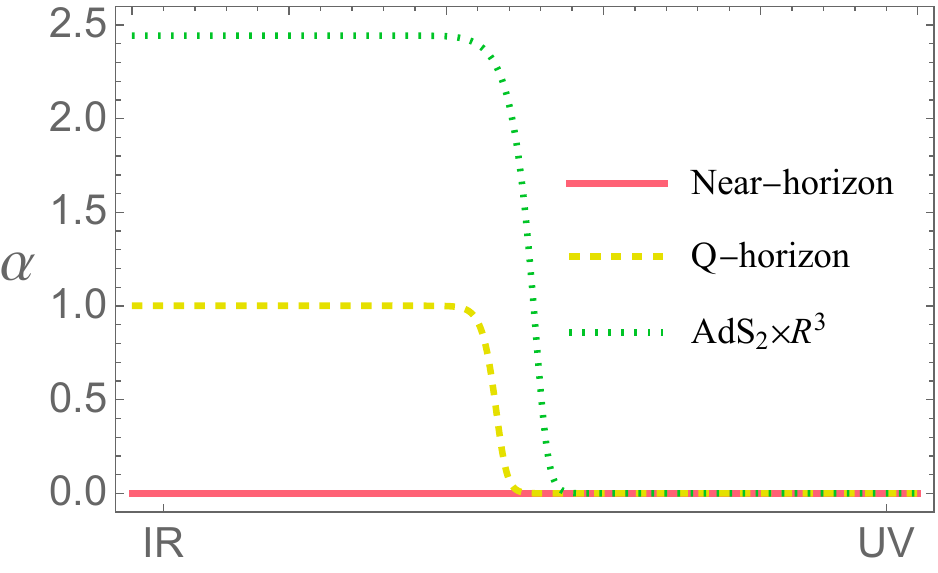}
\caption{Examples for RG flows from AdS$_{n+2}$ to near-horizon (finite $T$, zero $q$), Q-horizon (finite $T$ and $q$) or AdS$_2\times$ R$^3$ (zero $T$, finite $q$). For simplicity, we fix $\beta=0$.}
\label{AdSuv}
\end{figure}

The first class of RG flows contains solutions which start from the AdS$_{n+2}$ UV fixed point, and do not exhibit intermediate scaling regimes. Switching on a relevant deformation induced by heat, electric charge or their combinations, the solution can flow to near-horizon, Q-horizon or AdS$_2\times$ R$^n$ fixed points in the IR respectively. These RG flows can be found setting $\beta=0$, as shown in Fig.\ref{AdSuv}.

\begin{figure}[h]
\centering
\includegraphics[width=0.3\textwidth]{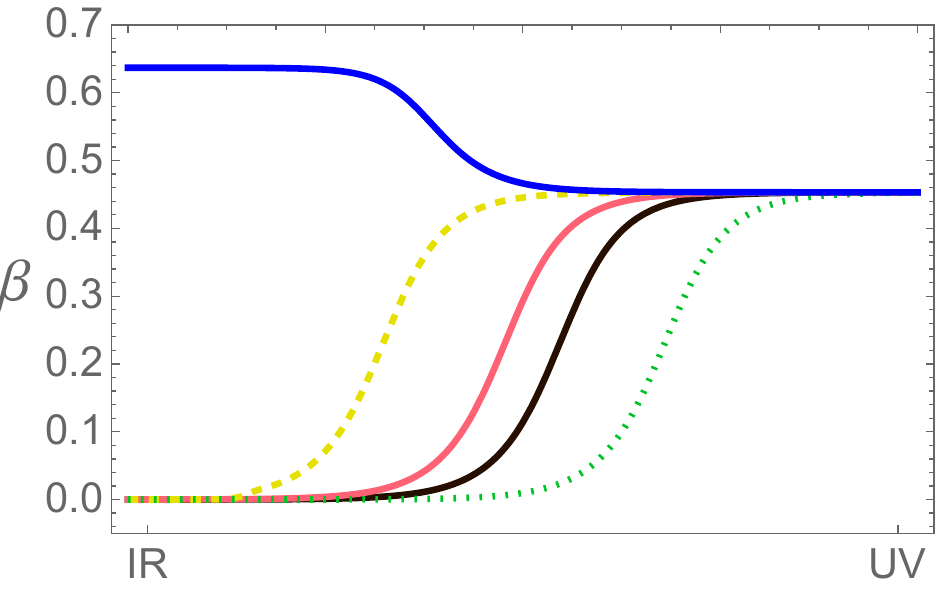}
\hspace{0.3cm}
\includegraphics[width=0.3\textwidth]{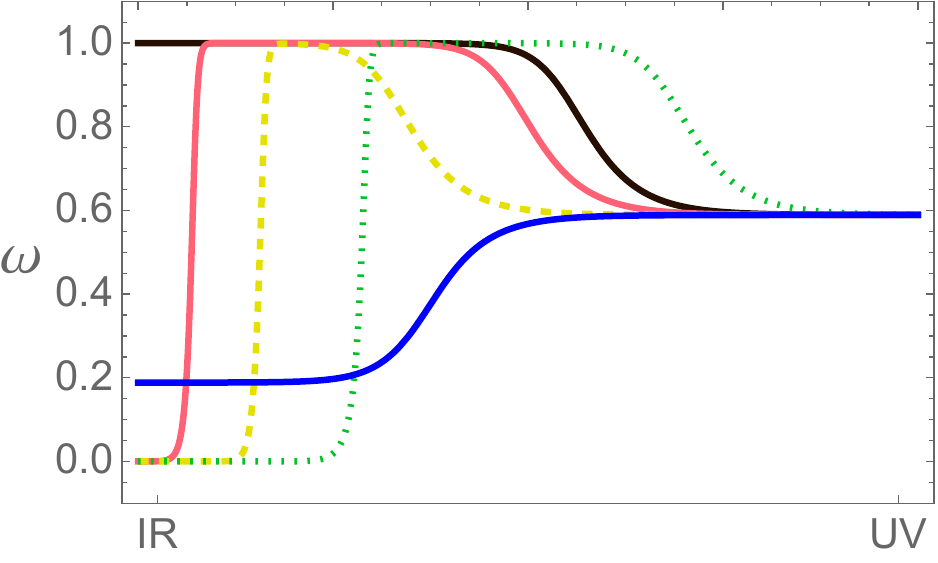}
\hspace{0.3cm}
\includegraphics[width=0.3\textwidth]{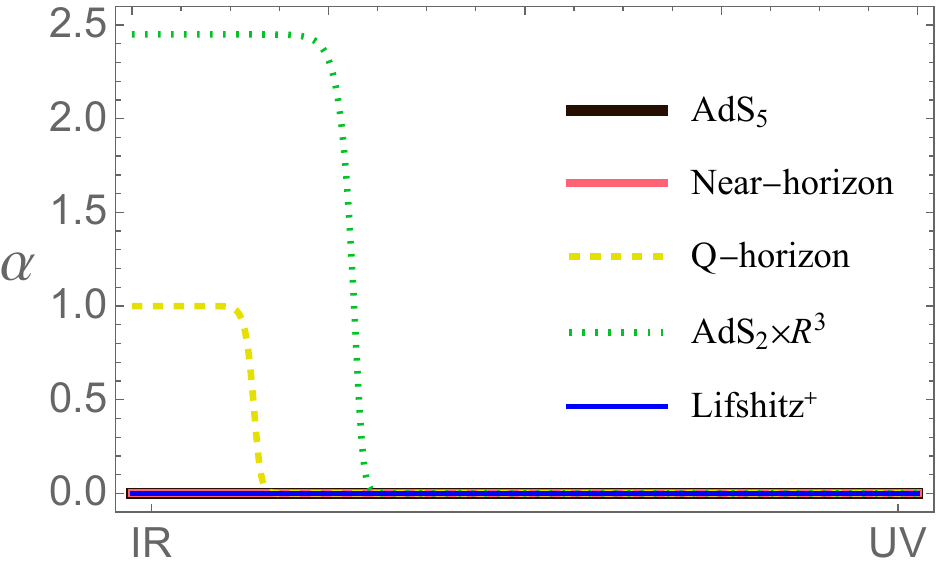}
\caption{Examples of RG flows from Lifshitz$^-$ to all the five possible IR fixed points with no intermediate scaling, realized with $m^2=4$.}
\label{Lifuv}
\end{figure}

When the time component of the Proca field $\beta$ is switched on, and simultaneously the mass is chosen properly that Lifshitz$^-$ branch of solutions exists, the non-vanishing $\beta$ deformation breaks relativistic symmetry. As a result, the UV fixed point can only be of the Lifshitz$^-$ type. As a second class, we show RG flows which start from a Lifshitz$^-$ solution and can flow to all the other possible fixed points in the IR, including AdS$_{n+2}$, near-horizon, Q-horizon, AdS$_2\times$R$^n$ and Lifshitz$^+$. These examples are illustrated in Fig.\ref{Lifuv}.

\begin{figure}[h]
\centering
\includegraphics[width=0.30\textwidth]{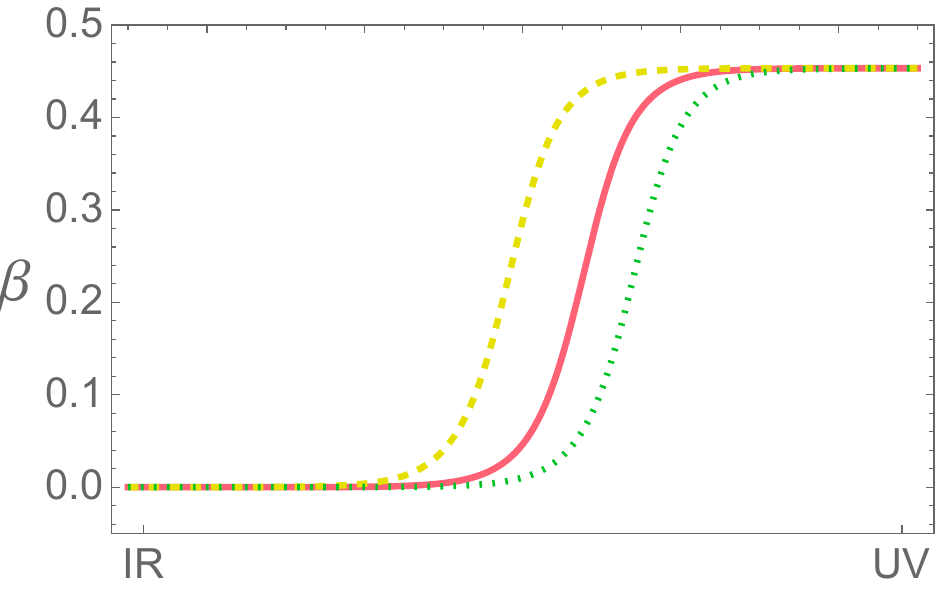}
\hspace{0.3cm}
\includegraphics[width=0.30\textwidth]{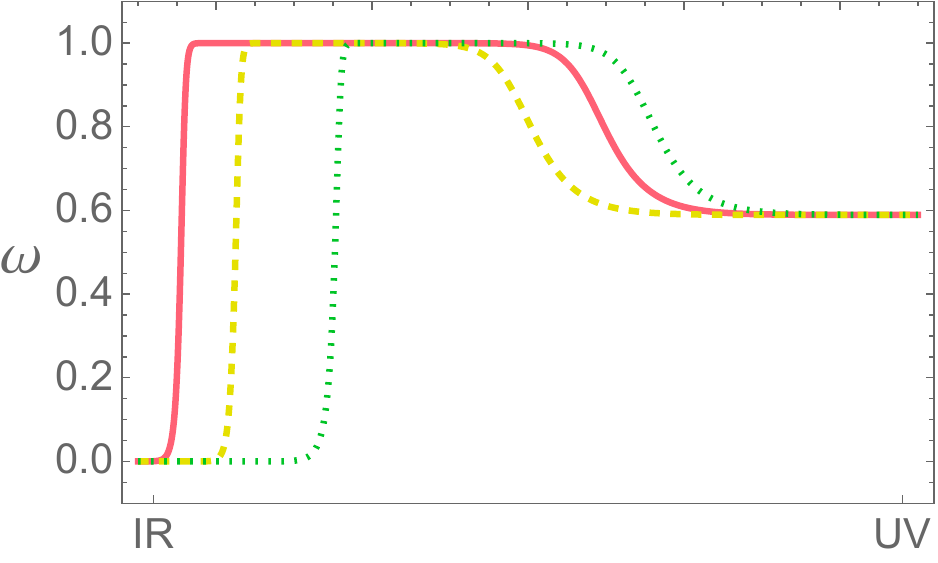}
\hspace{0.3cm}
\includegraphics[width=0.30\textwidth]{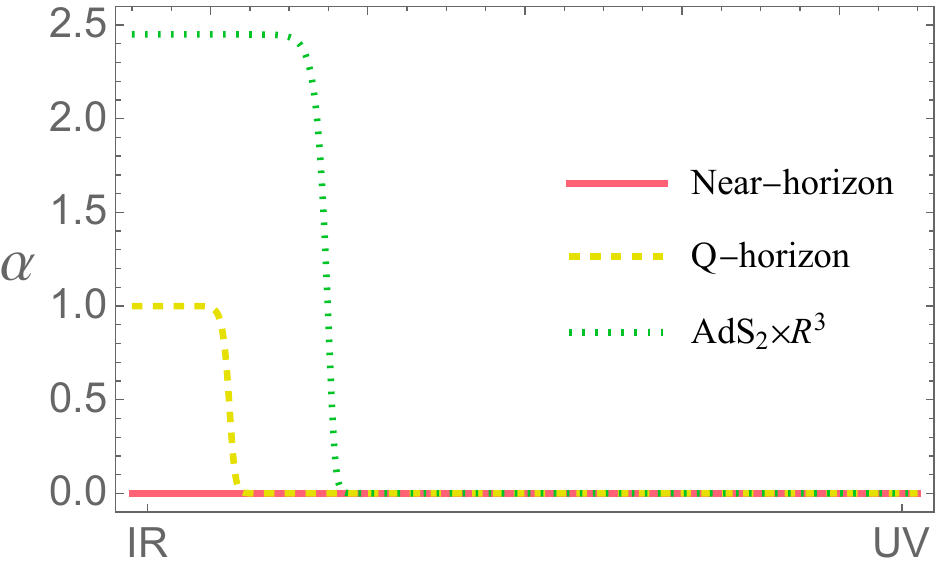}
\caption{Examples for RG flows from Lifshitz$^-$ to near-horizon, Q-horizon or AdS$_2\times$ R$^3$ with AdS$_{n+2}$ as an intermediate scaling region, realized with $m^2=4$.}
\label{LifuvAdSinter}
\end{figure}

The situation becomes more interesting when the RG flows have intermediate scaling regimes. From a field theoretical perspective, this is due to the CFT passing through a fixed point with both relevant and irrelevant directions. On the gravitational side, the intermediate scaling behaviors emerge via fine-tuning the initial conditions. In this third class, the RG flows still start from Lifshitz$^-$ to near-horizon, Q-horizon or AdS$_2\times$R$^n$ in the IR while display an AdS$_{n+2}$ intermediate scaling geometry, as shown in Fig.\ref{LifuvAdSinter}.

\begin{figure}[h]
\centering
\includegraphics[width=0.30\textwidth]{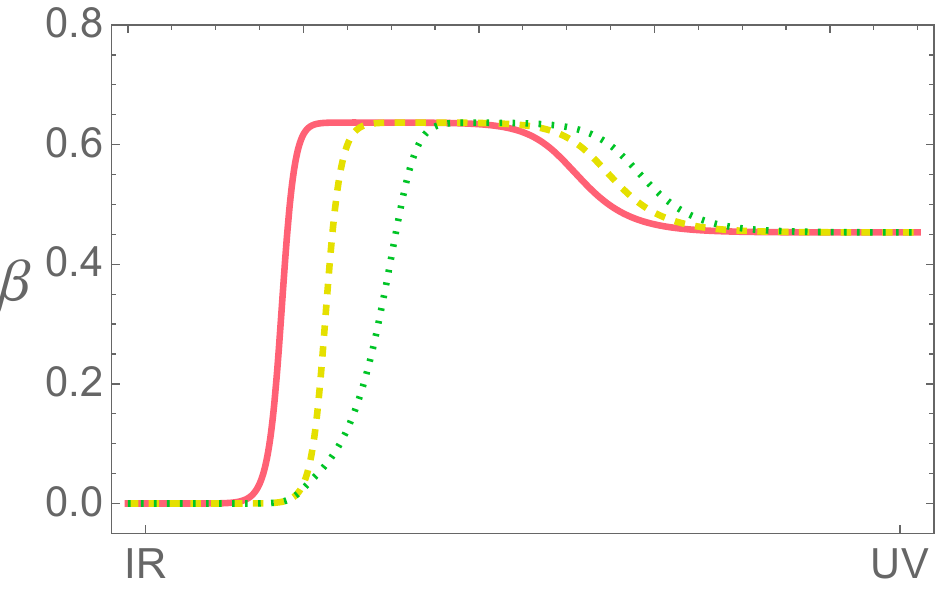}
\hspace{0.3cm}
\includegraphics[width=0.30\textwidth]{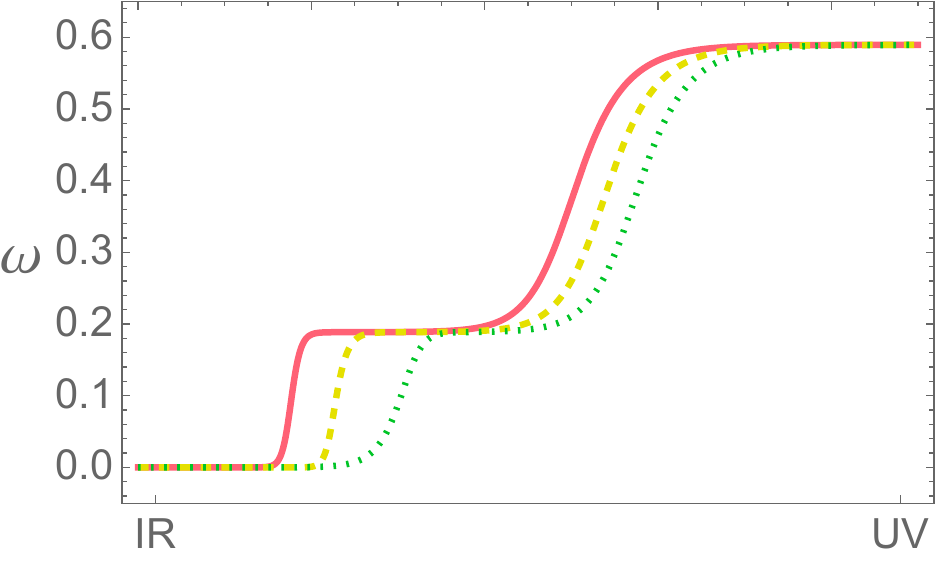}
\hspace{0.3cm}
\includegraphics[width=0.30\textwidth]{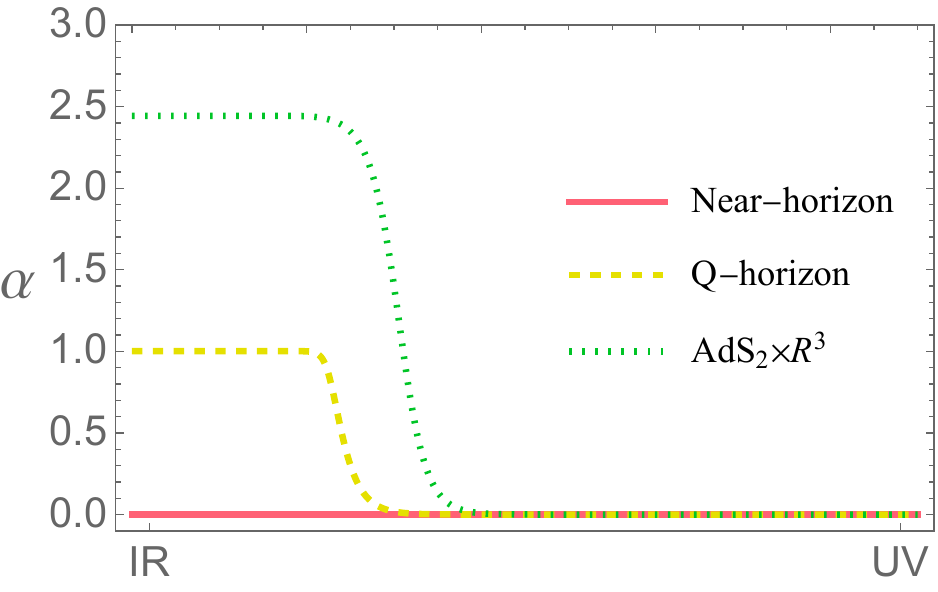}
\caption{Examples for RG flows from Lifshitz$^-$ to near-horizon, Q-horizon or AdS$_2\times$ R$^3$ with Lifshitz$^+$ as an intermediate scaling region, realized with $m^2=4$.}
\label{LifuvLifinter}
\end{figure}

The fourth class of RG flows exhibits properties similar to the third class, where the only difference is that the AdS$_{n+2}$ intermediate geometry is substituted by a Lifshitz$^+$. We show examples for this class in Fig.\ref{LifuvLifinter}.

\begin{figure}[h]
\centering
\includegraphics[width=0.30\textwidth]{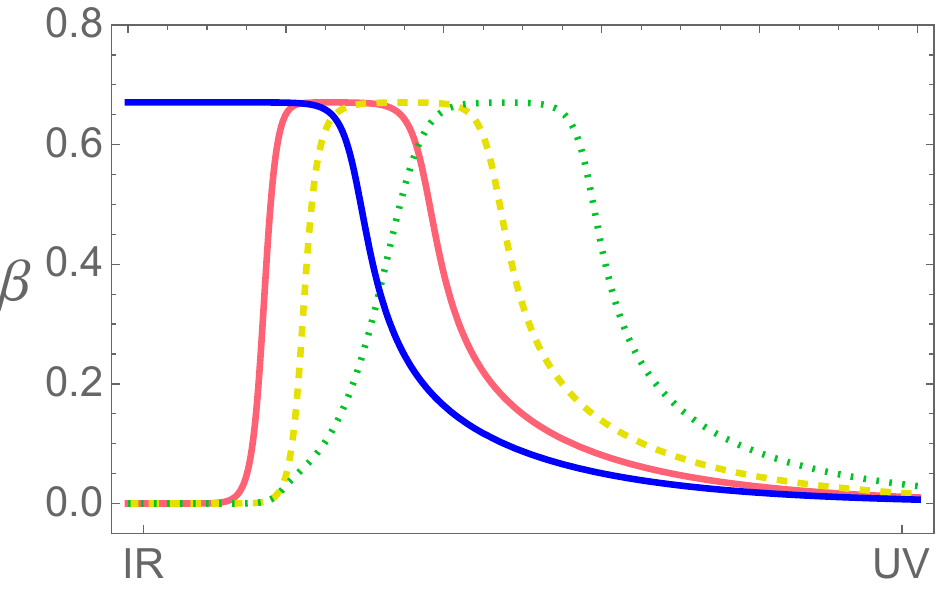}
\hspace{0.3cm}
\includegraphics[width=0.30\textwidth]{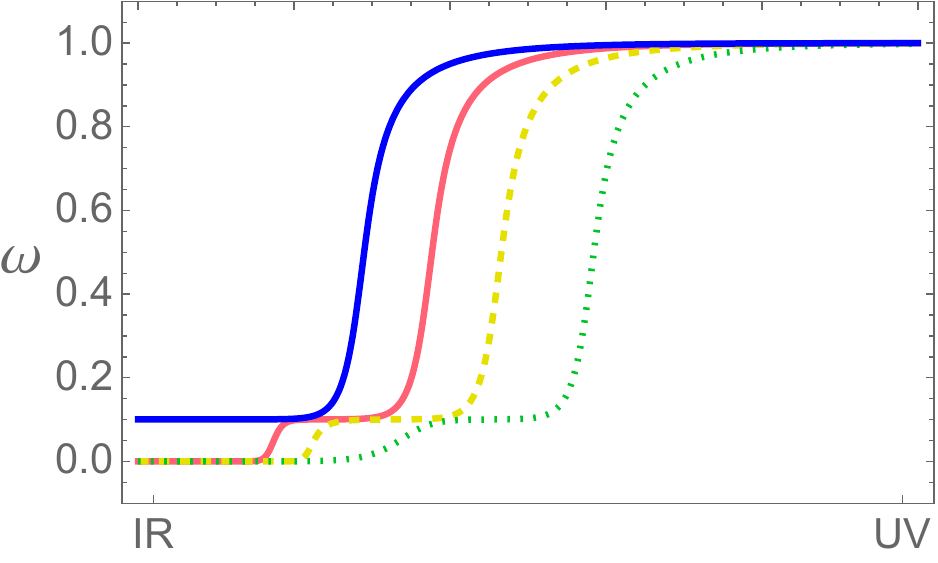}
\hspace{0.3cm}
\includegraphics[width=0.30\textwidth]{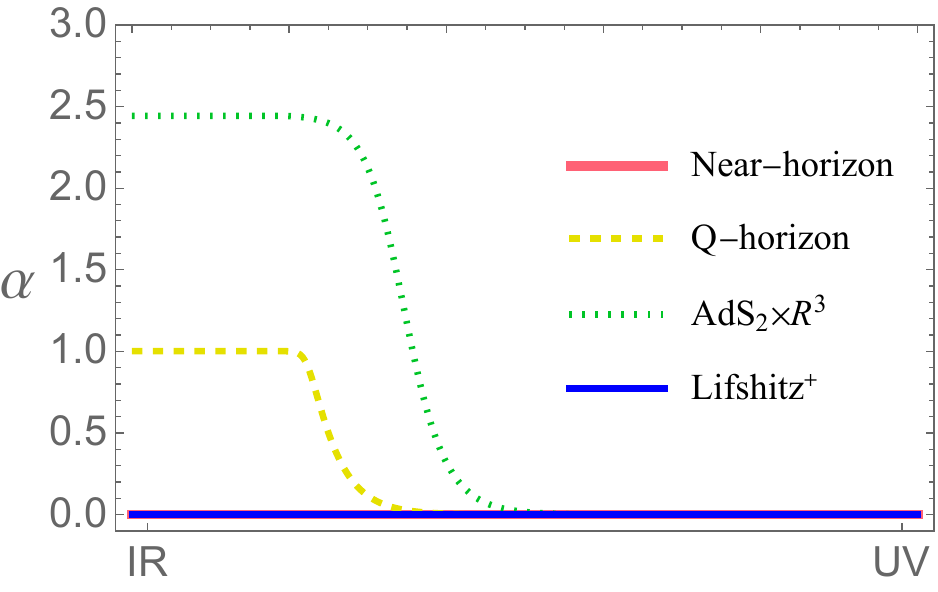}
\caption{Examples for RG flows from AdS$_{n+2}$ to near-horizon, Q-horizon or AdS$_2\times$ R$^3$ with Lifshitz$^+$ as an intermediate scaling region, realized with $m^2=2.8$.}
\label{AdSuvLifinter}
\end{figure}

Even though the finite Proca field breaks relativistic symmetry, the AdS$_{n+2}$ UV fixed point can indeed coexist with an infinitesimal Proca field as a relevant deformation. For example, this can be realized by choosing $m^2<n$ to make the Lifshitz$^-$ branch of solutions disappear. The corresponding possible RG flows are illustrated in the top right panel of Fig.\ref{fig:streams}. We define this as the fifth class, where the RG flows first suppress the electric charge to then vanish into the Lifshitz$^+$. Depending on the initial deformations, the solution can remain at the Lifshitz$^+$ fixed point, or flow further into near-horizon, Q-horizon or AdS$_2\times$ R$^3$ in the IR. We demonstrate this class in Fig.\ref{AdSuvLifinter}.


We summarize all the examples of RG flows discussed above in Table \ref{table:list}. 
\begin{table}[h!]
  \centering
  \begin{tabular}{c|ccc}
    \toprule
     \textbf{RG flows} & \textbf{IR} & \textbf{intermediate} & \textbf{UV} \\
    \bottomrule
     Fig.\ref{AdSuv}, Red &  Near-horizon & \slash & AdS \\[2mm]
     Fig.\ref{AdSuv}, Yellow &  Q-horizon & \slash & AdS \\[2mm]
     Fig.\ref{AdSuv}, Green &  AdS$_2\times$ R$^n$ & \slash & AdS \\[2mm]
     \midrule
     Fig.\ref{Lifuv}, Black &  AdS & \slash & Lifshitz$^-$ \\[2mm]
     Fig.\ref{Lifuv}, Red &  Near-horizon & \slash & Lifshitz$^-$ \\[2mm]
     Fig.\ref{Lifuv}, Yellow &  Q-horizon & \slash & Lifshitz$^-$ \\[2mm]
     Fig.\ref{Lifuv}, Green &  AdS$_2\times$ R$^n$ & \slash & Lifshitz$^-$ \\[2mm]
     Fig.\ref{Lifuv}, Blue &  Lifshitz$^+$ & \slash & Lifshitz$^-$ \\[2mm]
     \midrule
     Fig.\ref{LifuvAdSinter}, Red &  Near-horizon & AdS & Lifshitz$^-$ \\[2mm]
     Fig.\ref{LifuvAdSinter}, Yellow &  Q-horizon & AdS & Lifshitz$^-$ \\[2mm]
     Fig.\ref{LifuvAdSinter}, Green &  AdS$_2\times$ R$^n$ & AdS & Lifshitz$^-$ \\[2mm]
     \midrule
     Fig.\ref{LifuvLifinter}, Red &  Near-horizon & Lifshitz$^+$ & Lifshitz$^-$ \\[2mm]
     Fig.\ref{LifuvLifinter}, Yellow &  Q-horizon & Lifshitz$^+$ & Lifshitz$^-$ \\[2mm]
     Fig.\ref{LifuvLifinter}, Green &  AdS$_2\times$ R$^n$ & Lifshitz$^+$ & Lifshitz$^-$ \\[2mm]
     \midrule
     Fig.\ref{AdSuvLifinter}, Red &  Near-horizon & Lifshitz$^+$ & AdS \\[2mm]
     Fig.\ref{AdSuvLifinter}, Yellow &  Q-horizon & Lifshitz$^+$ & AdS \\[2mm]
     Fig.\ref{AdSuvLifinter}, Green &  AdS$_2\times$ R$^n$ & Lifshitz$^+$ & AdS \\[2mm]
     Fig.\ref{AdSuvLifinter}, Blue &  Lifshitz$^+$ & \slash & AdS \\[2mm]
       \bottomrule
  \end{tabular}
  \vspace{0.3cm}
      \caption{Full list of the RG flow geometries in our model. Lifshitz$^{+}$ refers to the upper branch of the Lifshitz solution while Lifshitz$^{-}$ to the lower one. All the listed flows are constructed numerically. {\bf Rows $1$-$3$}: AdS can flow to near-horizon geometry ($q=0, T>0$), Q-horizon ($q>0, T>0$) or AdS$_2$($q=q_e,T=0$). { \bf Rows $4$-$8$}: $\beta\neq 0$, $m^2$ admits Lifshitz$^{\pm}$, Lifshitz$^-$ can flow to five kinds of IR fixed points with no intermediate geometries. {\bf Rows $9$-$11$}: Similar to $5$-$7$, but with AdS intermediate geometry. { \bf Rows $12$-$14$}: Similar to $5$-$7$, but with Lifshitz$^+$ intermediate geometry. {\bf Rows $15$-$18$}: $m^2$ admits Lifshitz$^+$ only, AdS can flow to Lifshitz$+$ as an IR fixed point, or as an intermediate geometry and further into near-horizon geometry, Q-horizon or AdS$_2$.  }
      \label{table:list}
  \end{table}

\newpage

\section{Emergence of Lorentz symmetry in the IR}
\label{sec:emergence}
In Ref.\cite{Bednik:2013nxa}, it has been shown that in flows from a LV UV fixed point to a LI IR fixed point, the deviations of various physical observables from their relativistic form at low energies are suppressed by a power of the ratio between the infrared and ultraviolet scales. This power is controlled by the dimension $\Delta_\beta$ of the so-called least irrelevant Lorentz violating (LILVO) operator in the IR fixed point. 
More precisely, by the irrelevance of the LILVO, $\kappa\equiv \Delta_\beta-4$ (in a 3+1 dimensions).  
For instance, the difference between propagation speeds of different species scales like $\delta c \propto (E/\Lambda)^\kappa$ \cite{Bednik:2013nxa}.

Power-law running is  more efficient than the logarithmic running  common to weakly-coupled field theories.
Still, in the simplest models \cite{Braviner:2011kz,Bednik:2013nxa}
the irrelevance $\kappa$ turned out  to be not so large, $\kappa \lesssim 0.35$. This is equivalent to saying that the LV effects are still poorly suppressed and that the emergence of LI is not very efficient even at strong coupling. Here, our first aim is to consider a more general holographic framework and understand (I) what determines this power $\kappa$ and (II) how large that can be. As we will see, the introduction of a more general potential will allow us to strongly amplify the power-law and reach $\kappa \sim \mathcal{O}(1)$.

For the sake of illustration, we focus on the quadratic + quartic potential 
\bea
V(\zeta)\,=\,-n(n+1)+2\,m^2\,\zeta+\gamma\,\zeta^2\,
\eea
where $\gamma>0$ to avoid instabilities. In Sec.\ref{sec:fixed points}, we have derived the conditions that the fixed point solutions need to obey for this quadratic potential. When the Proca field vanishes, only the first term in the potential (the cosmological constant) determines the solutions. Therefore, the quadratic term only changes the properties of the Lifshitz fixed points where the Lifshitz exponent is determined from the Proca charge via $\beta_0=\sqrt{\frac{z-1}{2z}}$. 
As shown in Eq.\eqref{Lif2nd} and in Fig.\ref{lifFix}, as $\gamma$ increases, it enlarges the parameter region as a function $m^2$ that admits the existence of Lifshitz fixed points. 

\begin{figure}[h]
\centering
\includegraphics[width=0.65\textwidth]{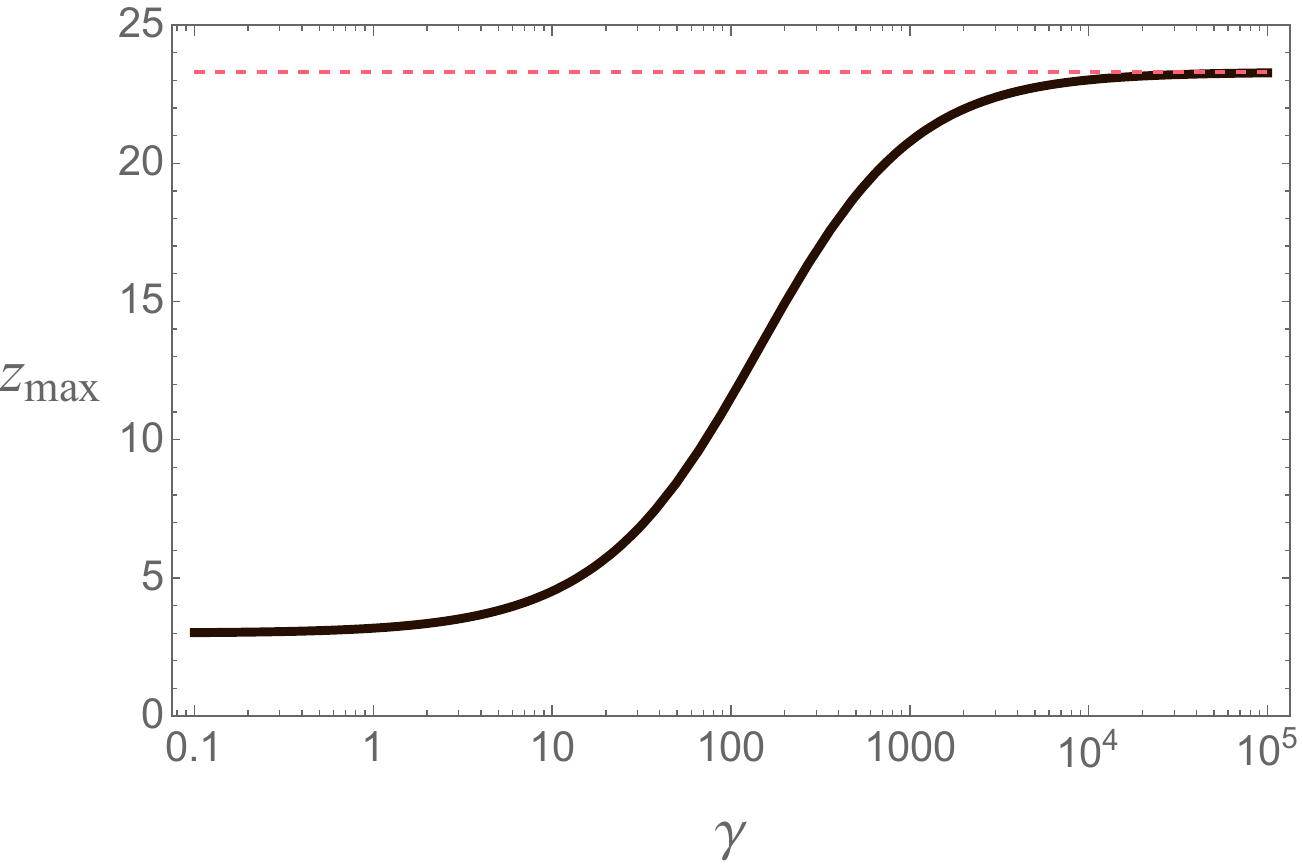}
\caption{TThe maximum dynamical exponent $z_{\text {max}}$ as a function of the quadratic coefficient $\gamma$, for $n=3$. The dashed red line indicates the upper bound $z_{\max}\approx 23.3$. }
\label{zmax}
\end{figure}

Moreover, the maximum value of the UV branch $z_{\text{max}}$ increases monotonically with respect to $\gamma$. As illustrated in Fig.\ref{zmax} for $n=3$, $z_{\text{max}}$ saturates an upper bound that is about $23.3$ for large values of $\gamma$. This upper bound on $z_{\text{max}}$ exists in general for $n\geq 3$, otherwise $z_{\text{max}}$ can increase with respect to $\gamma$ with no limitations. 

%%%%%%%%%%%%%%%%%%%%%%%%%%

We consider Lorentz violating deformations around the IR AdS$_{n+2}$ fixed points, which are encoded in the linearized solutions of the Proca field
\bea
\delta\beta(u)=c_1\,e^{-(\kappa+n+1)u}+c_2\,e^{\kappa u}
\eea
where 
\bea
\kappa \equiv \Delta_\beta-(n+1)
\eea
is the LILVO's \emph{irrelevance} in general dimensions. 

For the linear potential already studied in Refs.\cite{Braviner:2011kz,Bednik:2013nxa}, we can use the expression for $m^2$ in terms of $z$ and we get
\bea
\kappa=\frac{1}{2}\sqrt{1+n \left(\frac{4 (n+1) n z}{n^2+n z+(z-1) z}+n-2\right)}-\frac{n+1}{2},
\eea
where $z$ is the dynamical exponent of the Lifshitz fixed point in the UV, from which it flows into AdS in the IR. 

\begin{figure}[t]
\centering
\includegraphics[width=0.65\textwidth]{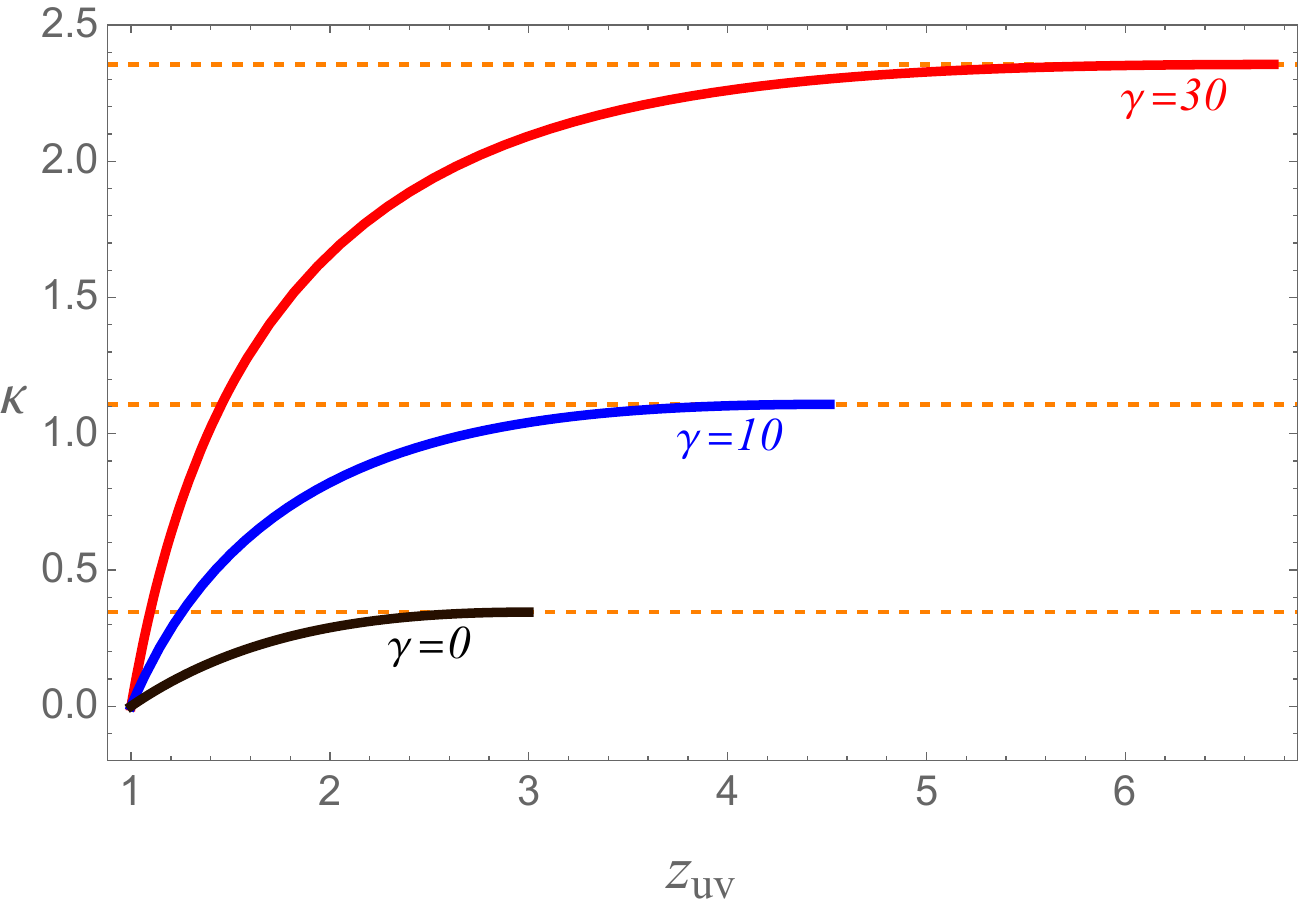}
\caption{ `Irrelevance' of the LILVO ($\kappa=\Delta_\beta-4$) as a function of the Lifshitz exponent $z_{UV}$, for different values of $\gamma$ for  $n=3$ ($3+1$ dimensional field theory). For the same $z_{UV}$, a larger $\gamma$ leads to a larger $\kappa$. This indicates that the quadratic potential makes it flow faster towards the LI attractive fixed points.}
\label{kappa}
\end{figure}

When the potential is generalized to the quadratic case in Eq.\eqref{quadrpot}, we obtain
\bea
\kappa=\sqrt{\frac{4 n^2 (n+1) z^2+\gamma  (z-1) \left(2 n^2+n (3 z-1)+2 (z-1) z\right)}{4z \left(n^2+n z+(z-1) z\right)}+\frac{(n-1)^2}{4}} \, -\frac{n+1}{2} \,.\nonumber
\eea
From this expression, we can find that  increasing the parameter $\gamma$ not only makes the maximum $z_{\text{max}}$ reachable in the UV larger but also enhances the dimension of the LILVO as well. The comparison between the linear case $\gamma=0$ and some finite $\gamma$ values in $n=3$ is shown in Fig.\ref{kappa}. 

There are three main outcomes from this exercise that appear clear in Fig.\ref{kappa}: 1) the irrelevance parameter $\kappa$ indeed can be extended to larger values by increasing the new coupling $\gamma$; 2) larger $\kappa$ is tied to larger values of $z_{UV}$;
3) moderately large values of $\gamma$, of order $10-20$ already accomplish to push the irrelevance $\kappa$ to an order one value, $1-2$. Such values of $\gamma$ do not introduce a hierarchically separated new strong coupling scale that affect the validity of the holographic model. Therefore, one concludes that it is possible to have RG flows with efficient emergence of LI in generic strongly coupled field theories.

\section{A Lifshitz Critical Line}
\label{sec:superpotential}

Critical lines (continuous line of fixed points) have been discussed mostly in condensed matter systems, see \textit{e.g.}, \cite{PhysRevB.76.134514,PhysRevLett.94.235702,PhysRevE.74.041124}. 
It is natural to ask whether it is possible to realize a critical line using holographic models.
In \cite{Braviner:2011kz,Bednik:2013nxa}, and all the other holographic models we are aware of, the Lifshitz fixed points that have been found are isolated points in the parameter space of the theory. Using our generalized framework, we can perform a systematic analysis to find the conditions for a holographic critical line. 

The first condition is that the field theory contains marginal couplings. For vector operators, this requires a \emph{massive} vector fields in AdS with mass $m^2=n$ (see \eqref{procadim}). A critical line (with different Lifshitz scaling at different points) can be obtained in case the operator is exactly marginal. Let us now see what this requirement translates into.

By looking at the algebraic equations for the Lifshitz fixed points, we can raise the following question. Can we find a special potential, which we will label the `Lifshitz Critical Line' potential $V_\lfp(\zeta)$, such that Lifshitz fixed points of any dynamical exponent $z$ exist? We remind the reader that we are using the notation
$$
\zeta\equiv B_\mu B^\mu~.
$$

On Lifshitz scaling solutions, 
Eq.\eqref{betalif} holds, which we can use to express the Lifshitz exponent $z$ as a function of $B_\mu B^\mu=-\beta_0^2$, \textit{i.e.},
\begin{equation}
z=\frac{1}{1-2\beta_0^2}=\frac{1}{1+2\zeta}\,,
\label{zbeta}
\end{equation}
where $0\leq \beta_0^2<1/2$. The equations of motion then reduce to
\begin{equation}\label{lfpeq}
\frac{2\, (2\, n\, \zeta+n)}{(2\, (n-1)\, \zeta+n) (2 \,n\, \zeta+n+1)}+\frac{V'_\lfp(\zeta)}{V_\lfp(\zeta)}\,=\,0\,.
\end{equation}
Clearly, having a potential that solves identically this simple differential equation will ensure that the Lifshitz solutions with free $\zeta$ exist. The physical boundary condition is $V_\lfp(0)=-n(n+1)$ such that when the Proca field vanishes, the AdS$_{n+2}$ solution is recovered. The solution for general $n$ then is
\begin{equation}
V_\lfp(\zeta).
\,=\,-n^{2+\frac{1}{n-1}} 
\frac{n+1\,+\,2\,n\, \zeta}{\left(n\,+\,2\, (n-1)\, \zeta \right)^{\frac{n}{n-1}} }
\,.
\label{superdef}
\end{equation}
Some comments are in order. First, we notice that there are no free parameters in \eqref{superdef}. It only depends on the field and the spatial dimension $n$. Together with the one-to-one correspondence between $\zeta$ and the Lifshitz exponent $z$ \eqref{zbeta}, this implies that the potential $V_\lfp(\zeta)$ leads to a (unique) continuous line of Lifshitz fixed points with $1 \leq z <\infty$.
Notice that the AdS$_{n+2}$ fixed point is attached to the continuous Lifshitz line as an endpoint. 

Expanding in powers of $\zeta$ one recognizes that indeed the mass term is $m^2=n$ as expected for a marginal vector operator. Note that for $n>2$ the potential has a branch cut starting at $\zeta=-n/2(n-1)$, however this is beyond the physical region, $\zeta > -1/2$ (leading to positive Lifshitz exponent). In fact,  $\zeta=1/2$ is a minimum (for any $n$) of the potential. For $-n/2(n-1) < \zeta < -1/2$ the  effective mass is tachyonic so it is expected to contain ghosts (see \textit{e.g.} \cite{Dvali:2007ks}), much in line with the would-be problematic values of $z<0$.

The solutions with $\zeta>-1/2$  instead are expected to be free from ghosts.  
The point $\zeta=-1/2$ (that would lead to $z\to\infty$,  like for the zero temperature Maxwell charged horizons) is expected to suffer from (infinite) strong coupling. Therefore, we expect the theory to be safe from ghosts and strong coupling (in the gravity side) so long as $\zeta$ is not too close to $-1/2$, from above. This can be confirmed by performing an analysis of perturbations around the solutions, which we leave for future work. For the present discussion it is enough to accept that the critical line exists as a weakly coupled theory in the gravity side up to some finite but large exponent $z_{max}\gg1$.

\begin{figure}[htbp]
\centering
\includegraphics[width=0.42\textwidth]{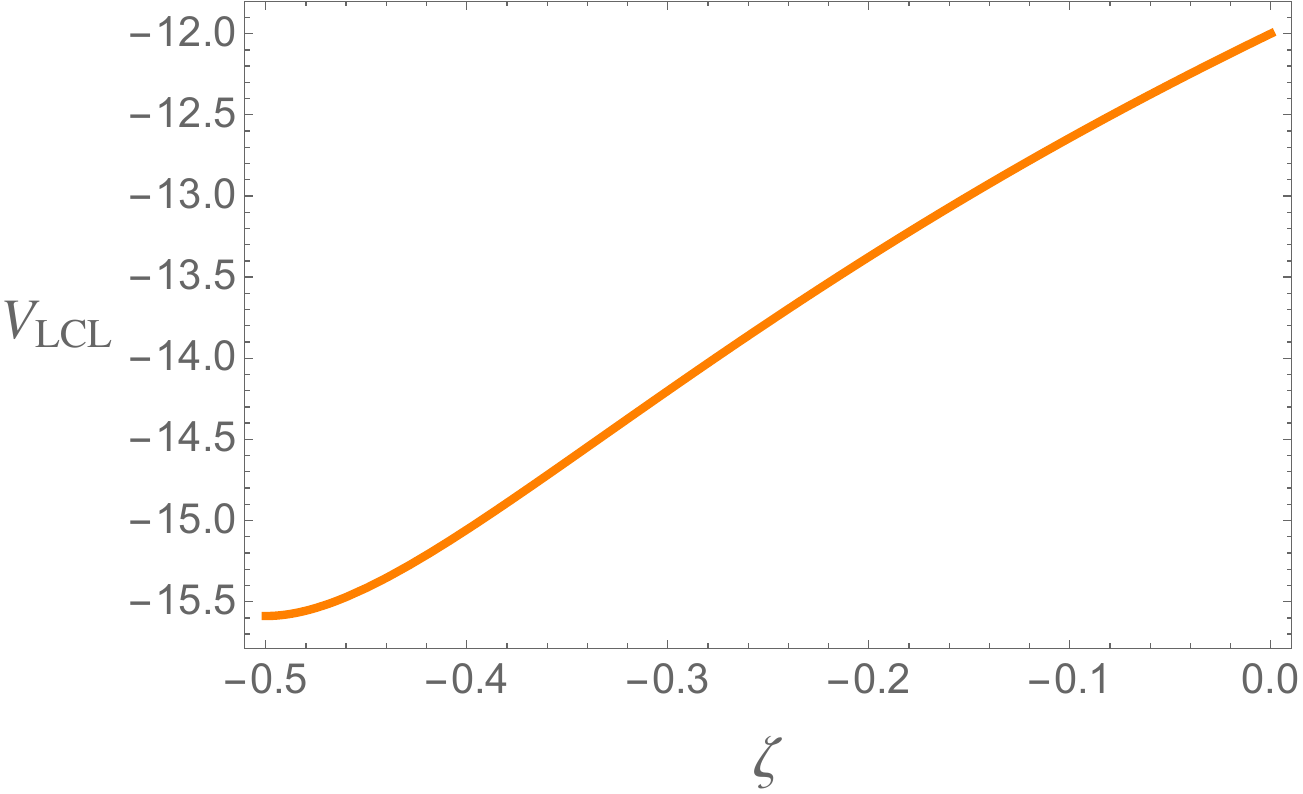}
\hspace{0.5cm}
\includegraphics[width=0.36\textwidth]{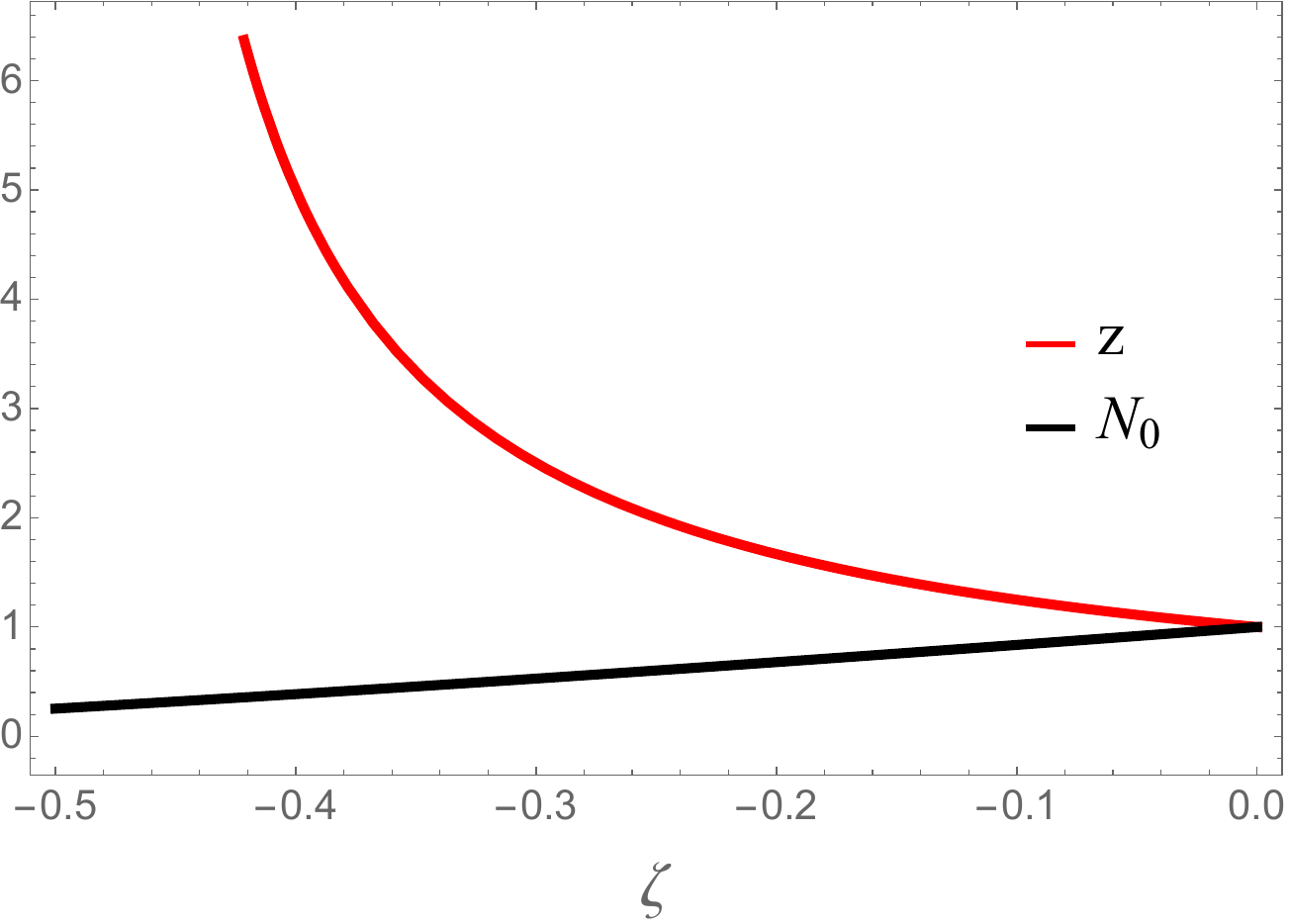}
\caption{The `Lifshitz Critical Line' $V_\lfp(\zeta)$ ({\bf left}) and the corresponding solutions of $z$ and $N_0$ ({\bf right}) for $n=3$. Here $\zeta=0$ corresponds to the AdS$_5$ case, while $\zeta=-1/2$ corresponds to $z=\infty$, \textit{i.e.}, AdS$_2 \times R^3$. }
\label{superpot}
\end{figure}

The critical line potential $V_\lfp(\zeta)$  is plotted in Fig.\ref{superpot}, together with the values of $z$ and $N_0$ in the fixed point solution. 

In sum, we have established that a continuous line of Lifshitz fixed points can be constructed by designing the potential to the special form \eqref{lfpeq}. This algorithm provides an efficient way to obtain all Lifshitz-type fixed points with $z\geq 1 $ up to large $z$ values while keeping the gravity side under control. Another important question is how tuned the potential \eqref{lfpeq} is, say,  under quantum corrections. We leave a proper analysis of this question for the future, however it seems relevant to note that there is a Lifshitz scaling symmetry (with field-dependent Lifshitz exponent) that certainly protects the form of the potential.

At this stage, another natural question arises. Do relevant or irrelevant directions around this line of Lifshitz fixed points exist? Or, equivalently, can these fixed points be the UV or IR endpoints of the RG flow? If the critical line can exist in the UV, do these fixed points flow to a relativistic IR fixed point and how? In this Section, we will answer these questions. 

\begin{figure}[h]
\centering
\includegraphics[width=0.45\textwidth]{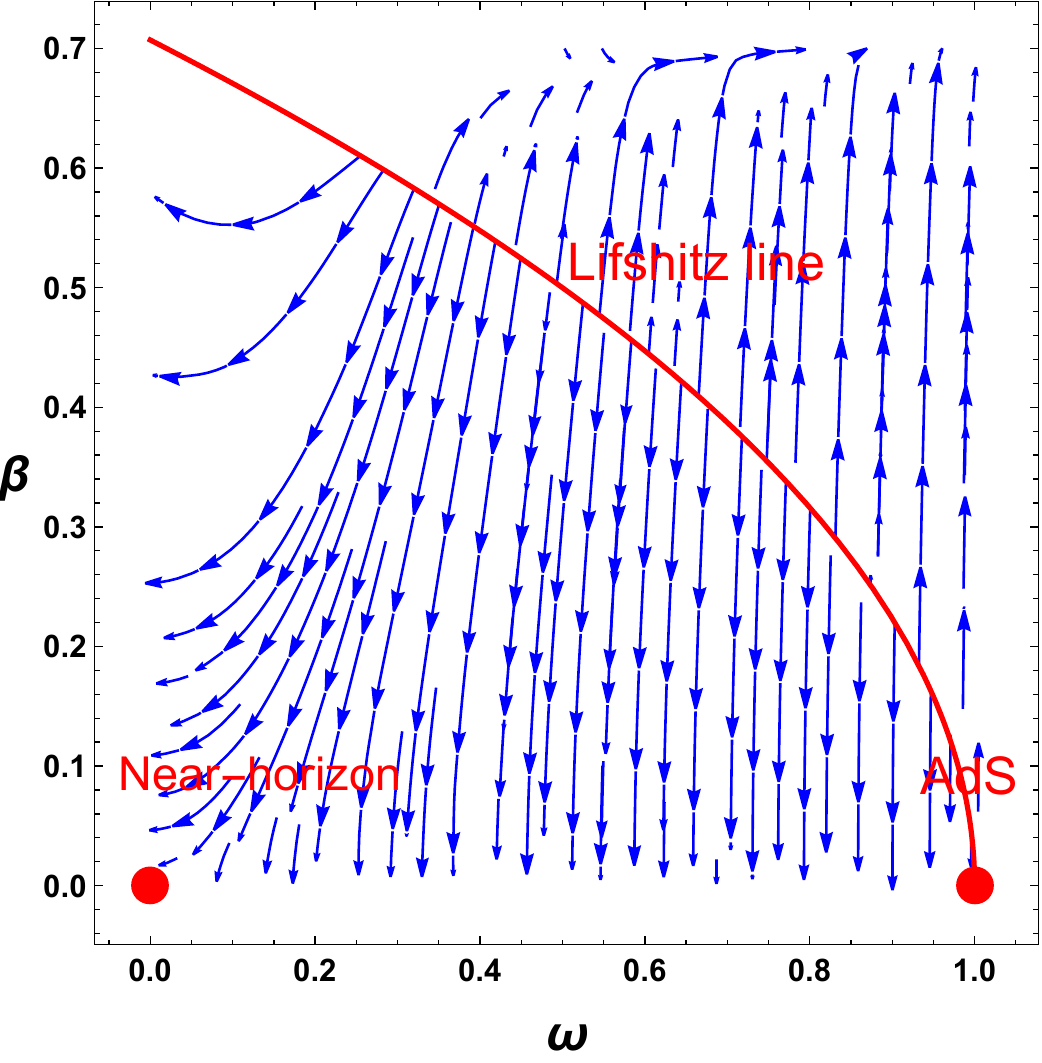}
\caption{Stream plot for the `Lifshitz Critical Line' case $V_\lfp(\zeta)$ for $n=3$ with $\alpha=0$. The Lifshitz fixed points form a continuous line with the AdS$_{n+2}$ lying at one of the endpoints corresponding to $V_\lfp(0)$. There are only relevant directions around the Lifshitz line, while only irrelevant directions around near-horizon, Q-horizon or AdS$_2$.}
\label{streamSupV}
\end{figure}

As discussed above, the relevant and irrelevant directions are encoded in the linear perturbations around the fixed points. 
Assuming  the line-potential, the other admitted fixed points/limits are the familiar near-horizon, Q-horizon and AdS$_2$ types. 
In the vicinity of the AdS$_{n+2}$ fixed point, the linear perturbations $\delta \alpha\sim e^{-nu}, \delta w\sim e^{-(n+1)u}$ and $\delta \beta \sim e^{-(n+1)u}$ provide the relevant directions so that the AdS$_{n+2}$ can be a UV fixed point. Similar arguments are valid near the Lifshitz fixed point, where the linear perturbations $\delta \alpha \sim e^{-nu/z}, \delta w\sim e^{-\frac{n+z}{z}u}$ and $\delta\beta\sim e^{-\frac{n+z}{z}u}$ induce the relevant directions. There are no irrelevant directions around the continuous Lifshitz line. As a result, the RG flows can only start from the Lifshitz line and end at the near-horizon, Q-horizon or the AdS$_2$ fixed points, which is demonstrated in the stream plot in Fig.\ref{streamSupV}. We notice that in the region $w\lesssim 0.15$, $\beta>0.3$ , the linearized approximation utilized in Fig.\ref{streamSupV} breaks down and one cannot visualize the RG flows on the two-dimensional plane anymore. This is the reason why the flow lines are missing in that region in Fig.\ref{streamSupV}.

\begin{comment}
Given this line-potential, the possible fixed points are the Rindler, charged Rindler and the AdS$_2\times$ R$^n$ for $\beta_0=0$, and the Lifshitz fixed point with $\beta_0=\sqrt{\frac{z-1}{2z}}$ that defines $z$. Via the linear perturbations around these fixed points as before, we obtain the following linear solutions 
\begin{itemize}
\item Near the AdS$_{n+2}$, $\delta \alpha\sim e^{-nu}, \delta w\sim e^{-(n+1)u}$ and $\delta \beta \sim e^{-(n+1)u}+b$. 
\item Near the charged Rindler horizon or AdS$_2\times$ R$^n$, $\delta \alpha\sim e^{2u}+a, \delta w\sim e^{2u}$ and $\delta \beta\sim b_1\,e^u+b_2\,e^{-u}$.
\item Near the Lifshitz fixed point, $\delta \alpha \sim e^{-nu/z}, \delta w\sim e^{-\frac{n+z}{z}u}+w_1$ and $\delta\beta\sim e^{-\frac{n+z}{z}u}+b_1$. 
\end{itemize}
Then we find that, the possible UV fixed point allowed by the line-potential is the Lifshitz fixed points together with the AdS$_{n+2}$ that form a continuous line with the exponent $1\leq z< \infty$. 
\end{comment}

\iffalse
\begin{figure}[h!]
\centering
\includegraphics[width=0.40\textwidth]{superVRindler1.pdf}
\includegraphics[width=0.40\textwidth]{superVRindler2.pdf}
\caption{Taking the line-potential $V_\lfp$ defined in Eq.\ref{superdef}, the RG flows can start from Lifshitz fixed points $1<z<\infty$ to near-horizon. }
\label{superVRindler}
\end{figure}
\fi

\begin{figure}[h!]
\centering
\includegraphics[width=0.30\textwidth]{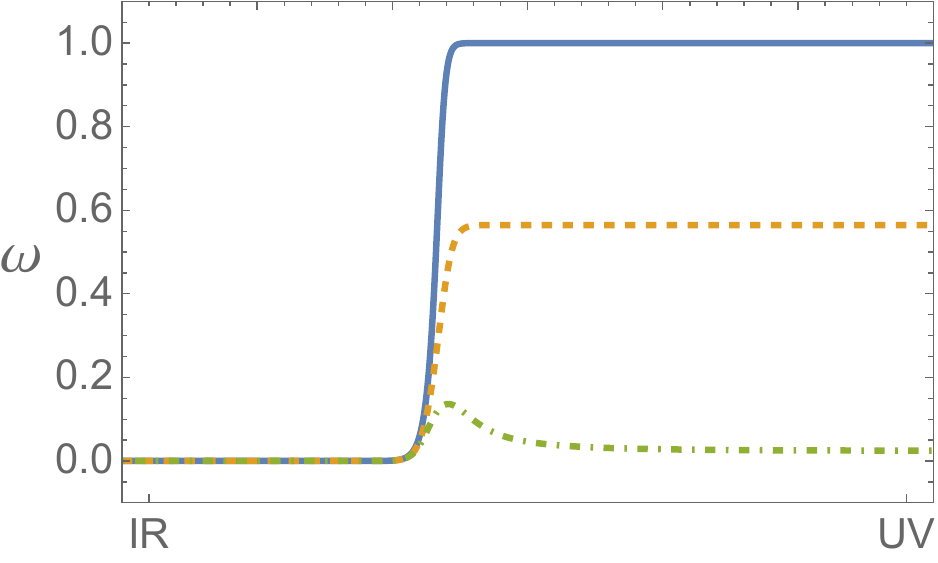}\hspace{0.3cm}
\includegraphics[width=0.30\textwidth]{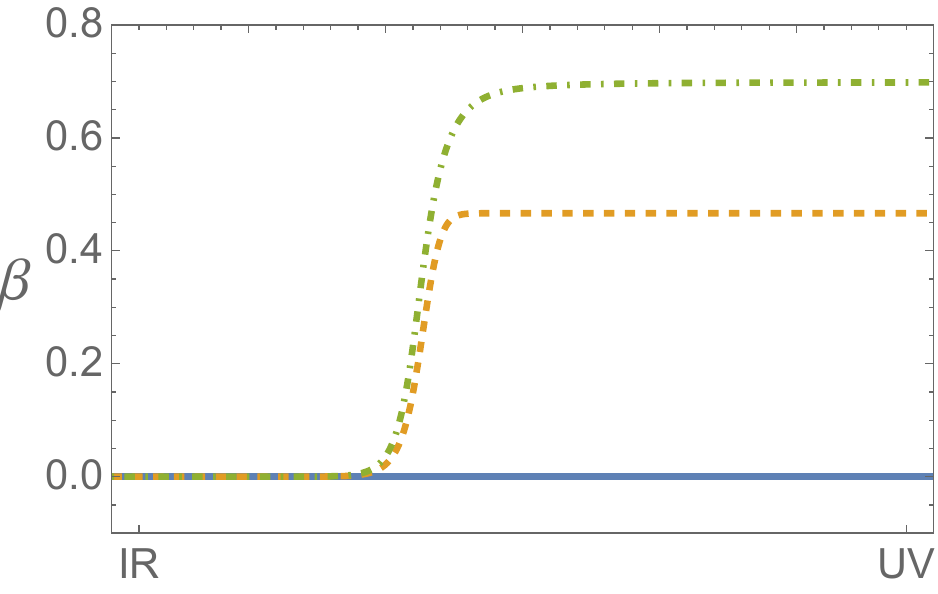}\hspace{0.3cm}
\includegraphics[width=0.30\textwidth]{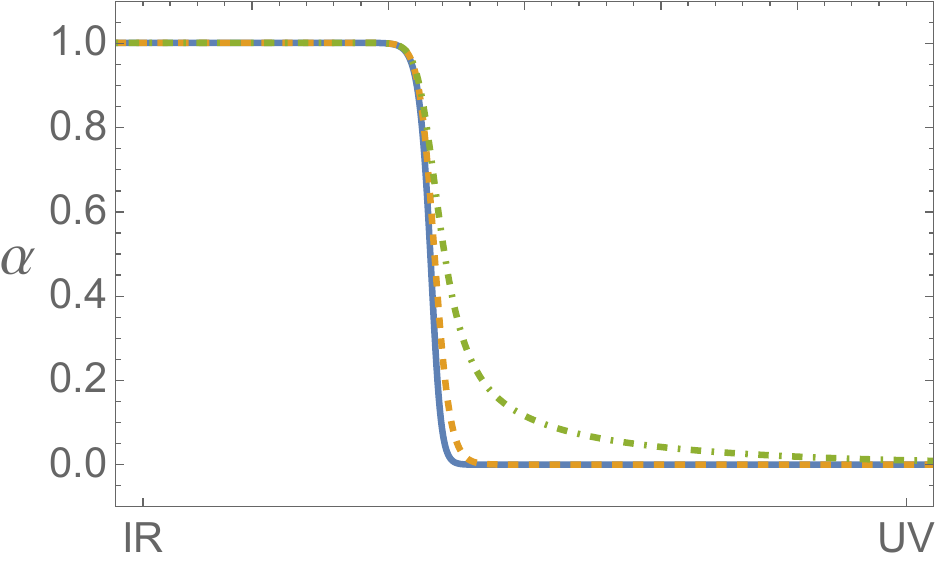}
\caption{Taking the `Lifshitz Critical Line' potential $V_\lfp$ defined in Eq.\ref{superdef}, the RG flows can flow from UV Lifshitz fixed points with arbitrary dynamical exponent, $1\leq z<\infty$. The respective dynamical exponent are: $z=1$ (blue), $z\simeq 1.76$ (orange), and $z\simeq 35.71$ (green). In this case, we have chosen the IR fixed point to be a Q-horizon geometry with $q=1$.}
\label{superVqRindler}
\end{figure}

Based on the instability analysis, we provide some examples of these RG flows. Via fine-tuning the irrelevant deformations around the IR fixed point, we find there exists no constraint on the value of Lifshitz exponent $z$ in UV, which is consistent with our previous analysis. These examples are illustrated in Fig.\ref{superVqRindler}. 

\section{Monotonicity of the $a$-function along the RG flows}
\label{sec:a-function}
In this Section, we investigate the monotonicity of the $a$-function for RG flows without LI, focusing in particular on the proposal of Ref.\cite{Caceres:2022hei}. 

We first briefly review the main idea in \cite{Caceres:2022hei}. In the so-called $n+2$ dimensional domain-wall ansatz
\bea
ds^2=e^{2A(\rho)}\left[-f(\rho)^2dt^2+e^{2\chi(\rho)}d\vec{x}^2_1+d\vec{x}_2^2\right]+d\rho^2\,,
\label{domain-wall}
\eea
where $t\in\mathbb{R}$, $\vec{x}_1\in\mathbb{R}^{\delta_1}$ and $\vec{x}_2\in\mathbb{R}^{\delta_2}$, with $\delta_1+\delta_2=n$ indicating the anisotropy in the spatial directions. $\rho\geq 0$ is the coordinate along the radial direction and $f(\rho)>0$ to satisfy Lorentzian signature. In the domain-wall ansatz, the $a$-function is proposed to be 
\bea
a(\rho)\sim e^{-\delta_1\chi(\rho)}\left[\frac{nf(\rho)}{\delta_1\left(A'(\rho)+\chi'(\rho)\right)+\delta_2A'(\rho)}\right]^n\,,  
\eea
and this function increases monotonically with respect to $\rho$ in the exterior of the black hole as constrained by the null energy conditions. 

In order to write the $a$-function in our notations, we compare the ``domain-wall" ansatz Eq.\eqref{domain-wall} and the ``RG-Gauge" ansatz Eq.\eqref{Ansatz} to find that\footnote{Here we assume $\delta_1=0, \delta_2=n$ for our isotropic system. We have checked that the opposite choice $\delta_2=0, \delta_1=n$ leads to the same results.}
\bea
d\rho=N(u)du\,,\quad f(\rho)=\frac{e^u}{e^{\int_0^uw(\xi)d\xi}}\,,\quad \text{and} \quad A'(\rho)=\frac{w(u)}{N(u)}\,. 
\eea
Then the $a$-function in the coordinates and ansatz of Eq.\eqref{Ansatz} is given by
\bea
a(u)\sim \left[\frac{e^u}{e^{\int_0^u w(\xi)d\xi}}\frac{N(u)}{w(u)}\right]^n\,.
\label{aofu}
\eea
Before jumping to the results, let us present some observations. First, even though the $a$-function is well-studied with the assumption of an asymptotic-to-AdS boundary, we will numerically verify its monotonicities with no such restrictions. Second, since $d\rho=N(u)du$ with $N(u)$ positive, the condition for the monotonicity of $a(u)$ should be $a'(u)\geq 0$. Furthermore, in the denominator of the r.h.s in Eq.\eqref{aofu}, the integral over the numerical solution usually results in obvious numerical errors and the exponential function even makes it worse. Therefore, we define the monotonicity function $m(u)$ from the derivative of the $a$-function as
\bea
\begin{split}
\label{eq:m(u)}
a'(u)&=\frac{nN^{n-1}}{w^{n+1}}\left(\frac{e^u}{e^{\int_0^uw(\xi)d\xi}}\right)^n\left(wN'+N(w-w^2-w')\right)\,\\
&\equiv \frac{nN^{n-1}}{w^{n+1}}\left(\frac{e^u}{e^{\int_0^uw(\xi)d\xi}}\right)^n m(u)\,,
\end{split}
\eea
where $N$ and $w$ are functions of $u$. Since $N,w$ and the exponential functions are always positive, then we only need to show that $m(u)\geq 0$ to confirm the monotonicity of $a(u)$. 

In the following, we numerically demonstrate the monotonicity of this $a$-function along several of the RG flows constructed in the previous Sections. The top left, top right, bottom left and bottom right correspond to the RG flows shown in Fig.\ref{Lifuv}, \ref{LifuvAdSinter}, \ref{LifuvLifinter} and \ref{AdSuvLifinter} respectively. Following the previous notations, different colors represent different IR fixed points. 

\begin{figure}[h!]
\centering
\includegraphics[width=0.45\textwidth]{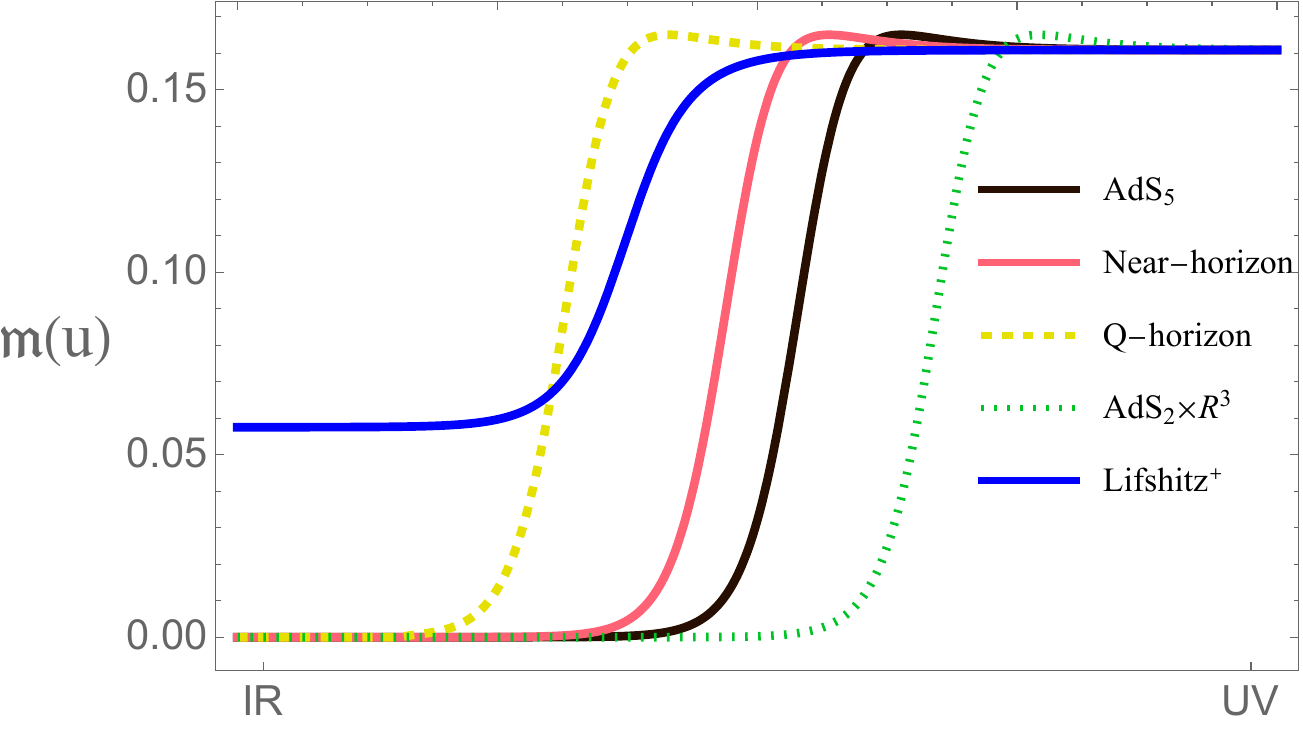}
\hspace{0.3cm}
\includegraphics[width=0.45\textwidth]{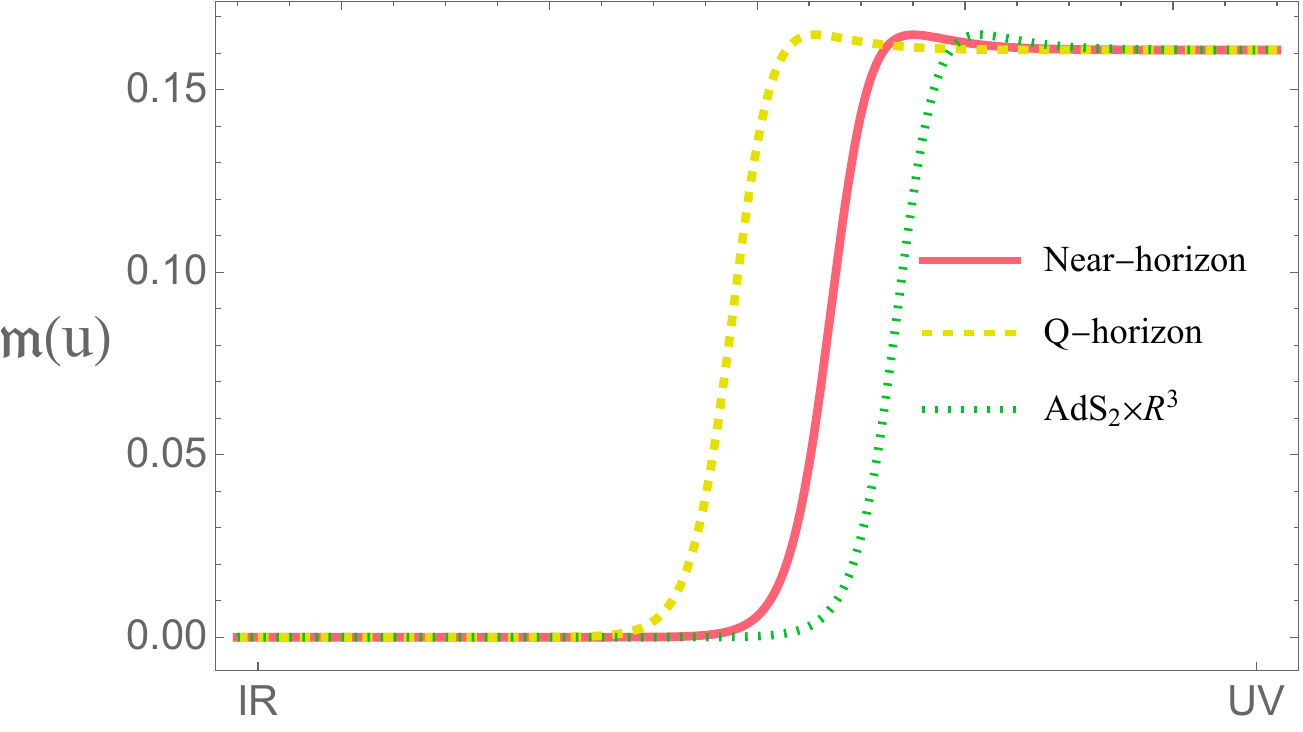}\\
\vspace{0.5cm}
\includegraphics[width=0.45\textwidth]{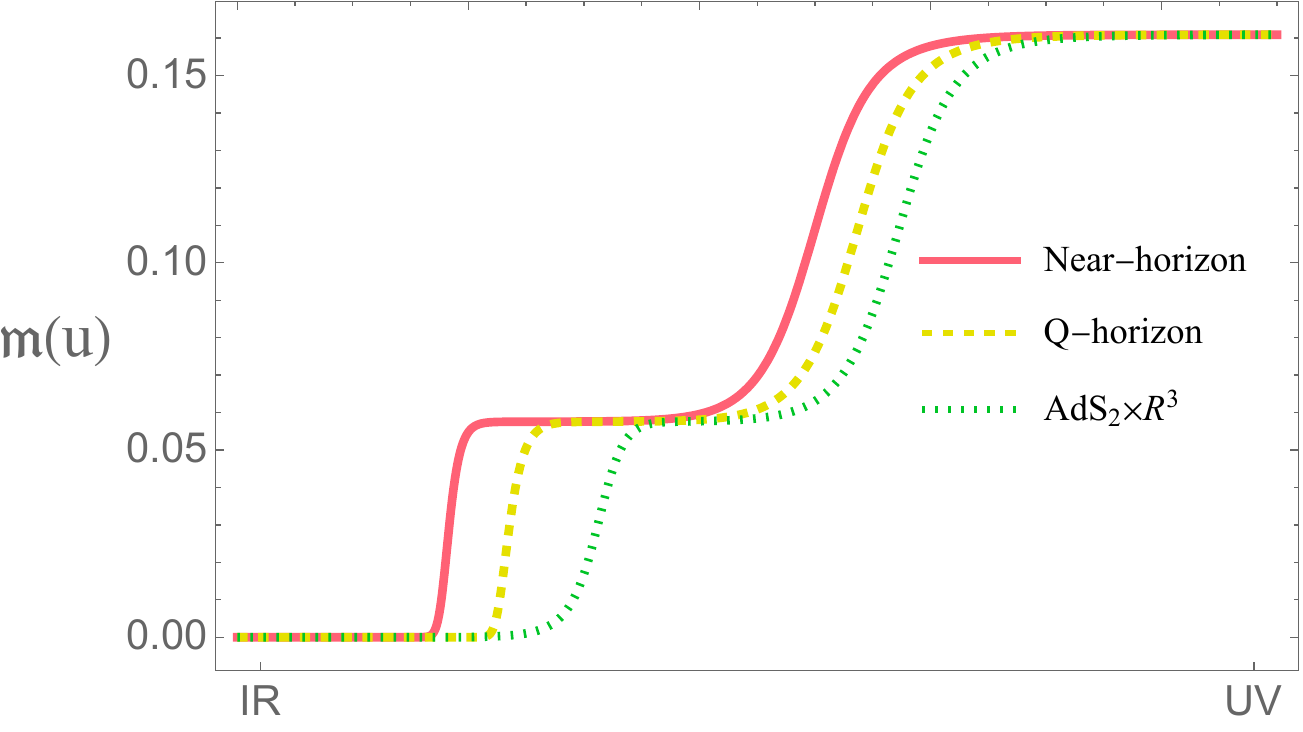}
\hspace{0.3cm}
\includegraphics[width=0.45\textwidth]{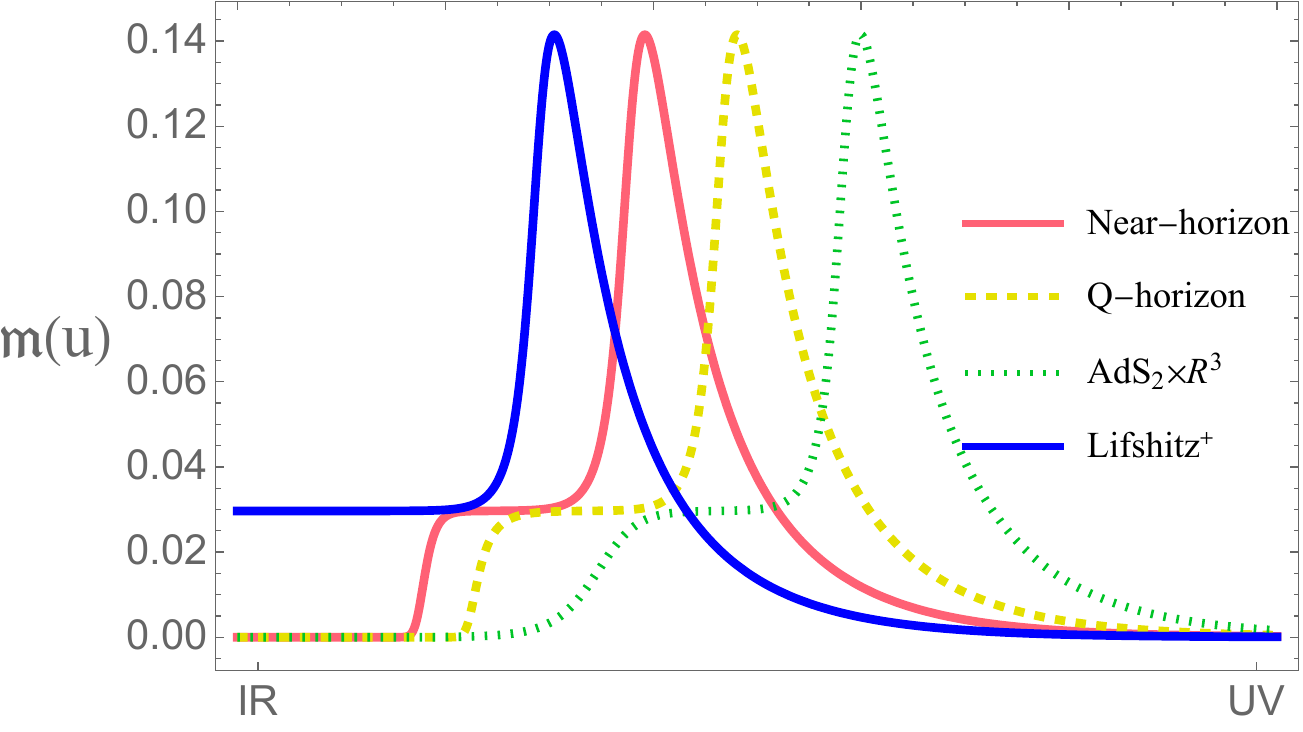}
\caption{A numerical proof for the monotonicity of the a-function proposed in \cite{Caceres:2022hei}. The monotonicity function $m(u)$ defined in Eq.\eqref{eq:m(u)} is always positive along the RG flows.}
\label{mfunction}
\end{figure}

We outline the regular behaviors in the function $m(u)$, which come from the properties of the fixed points. From the top left panel in Fig.\ref{mfunction}, one can find that $m(u)$ vanishes on AdS$_{n+2}$, near-horizon, Q-horizon and AdS$_2$ fixed points, while it develops a finite value near the Lifshitz$^+$ type fixed point. This finite value depends on the mass of the Proca field, and is an intrinsic feature of the Lifshitz fixed point since it emerges not only in the IR limit but also across the intermediate scaling regimes. Finally, when the RG flows start from the Lifshitz$^-$ fixed point, $m(u)$ is also a mass-dependent constant. In summary, we confirm that the proposal of \cite{Caceres:2022hei} is solid and valid also in our class of LV RG flows at finite temperature and finite charge density. 

We emphasize that the monotonicity of the $a$-function proposed in \cite{Caceres:2022hei} is guaranteed from the null energy condition (NEC) \cite{Curiel:2014zba}, which reads $T_{\mu\nu}k^\mu k^\nu \geq 0$
where $T_{\mu\nu}$ is the stress-energy tensor and $k^\mu$ an arbitrary null vector field.
In our case, this reduces to the condition on the potential $V(\zeta)$,
\bea
V'(\zeta)\geq  0\,,
\eea
which is recognized as the ghost-free condition discussed in Sec.~\ref{sec:superpotential}.
Since Einstein's equations imply
\bea
V'(\zeta)=n\,\frac{m(u)}{N(u)^3 \beta(u)^2}
\eea
and, by construction, $N(u)>0$, then the NEC is equivalent to the condition $m(u)>0$. As derived in Eq.\eqref{eq:m(u)}, this is the same requirement that ensures the monotonicity of the $a$-function, proving the initial statement. In other words, all the flows considered in our work respect the NEC, hence, they present a monotonic $a$-function, as proposed by \cite{Caceres:2022hei}.

\section{Discussion}
\label{sec:conclusion}

In this work, we have constructed a large class of holographic RG flows interpolating between LI and LV Lifshitz fixed points at finite temperature and finite charge density, and also involving intermediate scaling regions.
In order to do that, we have considered an Einstein-Maxwell-Proca model in arbitrary dimensions where the profile of the massive Proca field breaks explicitly Lorentz boosts. 
After classifying all the fixed points, we have analytically established their linear stability and constructed numerically a large class of holographic RG flows connecting them. 

These new geometric solutions could be of direct interest for applications of the holographic correspondence, especially to condensed matter systems where Lorentz symmetry is naturally broken. Nevertheless, our main motivation to construct such geometries was to understand better strongly coupled field theories without Lorentz invariance and in particular to address three related concrete questions.

First, as shown in \cite{Bednik:2013nxa}, strong coupling could accelerate the emergence of LI in the IR after adding an LV UV deformation. Nevertheless, at least in the models considered in \cite{Bednik:2013nxa}, such an enhancement was quite moderate. By using a more general class of holographic models without LI, we have seen that the emergence of LI can be more efficient. In more technical terms, the power $\kappa$ controlling the suppression of the LV effects along the RG flows can easily reach $\mathcal{O}(1-2)$ values, while in previous setups was limited to $\kappa \leq 0.35$, \textit{i.e.} too small. Holographically, this enhancement can be achieved by introducing quite minimal operators that are allowed, like simply a quartic term in the field.

Interestingly, the efficiency in the restoration of LI in the IR (parametrized by the dimension of the least irrelevant Lorentz violating operator) directly correlates with the Lifshitz exponent $z_\text{UV}$  achieved in the UV fixed point. In previous setups, the maximum  $z_\text{UV}$ was equal to the number of spacetime dimensions of the field theory. In our case, by deforming the holographic potential, such a number can be consistently larger. 

Thus, our holographic results suggest a surprising correlation between the violation of LI in the UV and the speed at which LI re-emerges in the IR. The more the scaling symmetry of the UV theory deviates from the LI case, \textit{i.e.}, the larger $z_\text{UV}$, the easier the LV effects are concealed in the IR. This conclusion deserves further investigation, especially in order to verify whether it is a universal result or a coincidence of the concrete class of holographic models considered here.

Taking advantage of the generalize Proca potential in the holographic model, we have been able to construct a critical line of Lifshitz fixed points. We are not aware of the existence of a critical line of LV fixed points in other holographic models. Interestingly, this implies that the UV geometry can get quite close to AdS$_2$, which corresponds to $z=\infty$. 

Our Lifshitz line of fixed points can be used as the basis to implement a technically natural spontaneous breaking of scale symmetry, in a setup without LI. By the same arguments of \cite{CPR} (see also \cite{Bellazzini:2013fga,Coradeschi:2013gda} and \cite{Megias:2014iwa,Pomarol:2019aae}), a small deformation of the critical line must generate an infrared scale (\textit{e.g.}, confinement) together with a naturally light dilaton, understood as the pseudo-Goldstone boson of spontaneously broken scale symmetry. 
One can even envision the situation where this happens concurrently with LI being emergent in the IR. In that case, though, our arguments suggest that the emergence of LI would be slow -- controlled by a marginal operator.

Finally, we have utilized the constructed holographic RG flows to test the a-function recently proposed in \cite{Caceres:2022hei}. We have confirmed the monotonicity of such a function along our RG flows, consistent with the validity of the null energy condition for the gravitational solutions considered in our work.

In the future, it would be interesting to extend our holographic RG flows to scenarios with even less symmetries, breaking for example rotational \cite{Cremonini:2014pca} and translational invariance \cite{Baggioli:2021xuv,RevModPhys.95.011001}. This could lead to potential applications to condensed matter systems and to understand better the breaking and emergence of spacetime symmetries in strongly-coupled field theories. So far, we have not investigated any transport properties of the dual field theories. It would be interesting to pursue this route and for example study the behavior of the shear viscosity and of thermoelectric conductivities on these RG flow solutions. In particular, it could be beneficial to understand in more detail the effects of the emergent scaling symmetries at different energy scales and how those are reflected in measurable observables. Moreover, it would be interesting to test with our solutions other holographic c-function proposed using entanglement entropy (for example that in \cite{Chu:2019uoh}). Finally, it would be fruitful to identify more stringent consistency constraints (in addition to the one imposed by the NEC, $V'(\zeta)\geq 0$) on the Proca model considered in this work. Studying the linear stability of perturbations (\textit{e.g.}, quasinormal modes) around these backgrounds is a promising method to achieve this task.

We leave these questions for future studies.

\section*{Acknowledgements}
We thank D.Giataganas and S.Cremonini for enjoyable discussions and useful comments about the subject of this paper. We thank Maria Vozmediano for explanations about the strongly coupled nature of graphene. X.W. and M.B. acknowledge the support of the Shanghai Municipal Science and Technology Major Project (Grant No.2019SHZDZX01) and the sponsorship from the Yangyang Development Fund. OP acknowledges the support from the Departament de Recerca i Universitats from Generalitat de Catalunya to the Grup de Recerca ‘Grup de Física Teòrica UAB/IFAE’ (2021 SGR 00649) and the Spanish Ministry of Science and Innovation (PID2020-115845GB-I00/AEI/10.13039/501100011033). IFAE is partially funded by the CERCA program of the Generalitat de Catalunya.

\appendix
\section{Equations of motions and fixed points}
\label{app:A}
Using the RG-Gauge ansatz \eqref{Ansatz}, the equations of motion for the background reduce to a second order differential equation for $\beta(u)$, two first order differential equations for $w(u)$ and $\alpha(u)$, and finally two algebric constraints for $N(u)$ and $N'(u)$.\\
The full system of equations reads:
\begin{align}
&\,N^2\,=\,\frac{\kappa}{U}\,,\nonumber\\[0.1cm]
&\,N'^2\,=\,\frac{-\kappa^2 \beta ^2 (n-1) V'+n\, V (-2 \,\rho^2+w 	\,(\tau+n)+1)+2 \,\alpha ^2 \,n\, (w \,(\,(n-1)^2 w+3\,
   n-2)+1))^2}{n^2 U^3}\,,\nonumber\\[0.1cm]
   &\,\alpha '+\alpha \, n \,w\,=\,0\,,\nonumber \\[0.2cm]
   &\,n\,U\,w'\,=\,((n-1) w+1) \left((n-1) \,n\, w^2 \left(2 \,\alpha ^2+\beta ^2 V'\right)+2 \,n\, w \left(2 \,\alpha ^2+\beta ^2\, V'\right)+2\, \beta ^2 \,\rho^2 \,V'\right)+\nonumber\\&+n \,w\, V \left(w \,(-\,n\,w\,+n+w-3)-2
   \left(\rho -1\right) \left(\rho +1\right)\right)\,,\nonumber\\[0.3cm]
   &\,\beta ''-\beta \, (2\, \beta\,  (n-1)\, \rho-n)\, V' (2\, \rho^2+n\, w\, ((n-1)\,
   w+2))+2\, n\, V\, (2 \,\beta ^3+2\, \beta '^3+6 \,\beta ^2\, \beta '+\nonumber\\&+\beta \, (6\, \beta '^2+w \,((n-1)
   \,w+2)-1)+w \,\beta ' ((n-1) \,w+2))-\nonumber\\&-4 \,\alpha ^2 \,n\, ((n-1)\, w\, \beta '\, ((n-1)\, w+2)+\beta \, ((n-1)\,  w+1)^2)\,=\,0\,,
   \label{GenEQs}
\end{align}
where we omitted all the u coordinate dependences and the argument of $V$ and $V'$ is always intended to be $B_\mu B^\mu\equiv\,-\beta^2$. We also define, because of space limitations, the following combinations: $\sigma=\beta+\beta'$, $\tau=w\,(1-n)-2$, $\kappa= n \,w\,\tau-2\,\sigma^2$, $U=2\,\alpha^2+V$. 
Fixed points are described by configurations for which all the ansatz functions take constant values $\{\alpha_0,N_0,w_0,\beta_0\}$. Introducing this ansatz in the system of equations \ref{GenEQs} we can get the equations of motion for the fixed points in full generality, without making explicit any form for the potential $V$.
All in all the equations for the fixed points read:
\begin{align}\label{FixP}
\begin{split}
&\,w_0\,\alpha_0\,=\,0\,,\\[0.1cm]
&\,N_0^2\,=\,\frac{n\,w_0 (-n \,w_0+w_0-2)-2 \,\beta_0^2}{2 \,\alpha_0^2+\hat{V}}\,,\\[0.2cm]
&\,\beta_0\, (\,-\,(2 \,\beta_0^2 \,(n-1)\,-\,n)\, (2\, \beta_0^2\,+\,n \,w_0\,(\,(n-1)\, w_0\,+\,2))\,
   \hat{V}'+\\&+2\, n\, \hat{V} (2 \,\beta_0^2\,+\,w_0\, (\,(n-1) \,w_0\,+\,2)\,-\,1)\,-\,4\,
   \alpha_0^2 \,n\, (\,(n-1)\, w_0\,+\,1)^2)\,=\,0\,,\\[0.25cm]
&\frac{1}{n \,(2 \,\alpha_0^2+\hat{V})}\,\Big[(\,(n-1) \,w_0+1) \,(\beta_0^2\, (2\,\beta_0^2+n\,w_0\, (\,(n-1)\,w_0+2))\, \hat{V}'+\\&+2\, \alpha_0^2\, n\,w_0\, ((n-1) \,w_0+2))+n \,w_0 \,\hat{V} \,(-2 \,\beta_0^2\,+\,w_0\,(-\,n\, w_0+\,n+\,w_0-\,3)\,+\,2)\Big]\,=\,0,\\[0.25cm]
&\,\frac{\sqrt{n \,w_0\, (-n\,w_0+w_0-2)-2 \beta_0^2}}{n \left(2 \,\alpha_0^2+\hat{V}\right)^{3/2}}\,\,(\beta_0^2\, (n-1) (2\, \beta_0^2+n \,w_0\, ((n-1) \,w_0+2)) \hat{V}'+\\&+n\,\hat{V} (-2\,\beta_0^2+w_0\, (n (-w_0)+n+w_0-2)+1)+2 \alpha_0^2 \,n\, (w_0 ((n-1)^2
   w_0+3\, n-2)+1))=0\,,
\end{split}
\end{align}
where $\hat{V}$ stands for $V(-\beta_0^2)$.

\section{Field theories with Lifshitz scaling}
\label{app:B}
In this Appendix, we derive the relation between the mass of a bulk vector field and the conformal dimension of the dual operator in theories with Lifshitz scaling, and show how that recovers the relativistic results when we set $z=1$. 

We consider a spacetime which enjoys the Lifshitz symmetry 
\bea
\vec{x}\rightarrow b\vec{x}\,,\quad t\rightarrow b^zt\,.
\eea
From dimensional analysis, we have $\left[k_i\right]=\left[\partial_i\right]=\left[E\right]=\left[\Lambda\right]=1$ with $E$ the energy scale and $\Lambda$ the UV cutoff. Then, we can conclude $\left[x^i\right]=-1$ for spatial directions, and $\left[\omega\right]=\left[\partial_t\right]=z$ with $\left[t\right]=-z$ for the time direction. 

Now, consider a term $S^{(0)}_i$ in the bare action defined at the energy scale $\Lambda_0$:
\bea
S^{(0)}\left[\Lambda_0\right]\supset S^{(0)}_i=\int dt\,d^{n}x~g^{(0)}_i\mathcal{O}_i\,,
\eea
where $\left[\mathcal O_i\right]=\Delta_i$, the bare coupling $g^{(0)}$ has mass dimension $\delta_i$ and can be reparametrized as $g^{(0)}=\lambda_i^{(0)}\Lambda_0^{\delta_i}$ via a dimensionless coupling $\lambda_i^{(0)}\sim O(1)$. Since the effective action is dimensionless, $\delta_i=n+z-\Delta_i$. Then, we assume that all the dimensional quantities are controlled by the energy scale $E$ we are interested in, that is 
\bea
k_i\sim E\,,\quad \omega\sim E^z\,,\quad \mathcal{O}_i\sim E^{\Delta_i}\,.
\eea
Then, we have 
\bea
\begin{split}
S_i\sim \int dt\,d^nx~g^{(0)}_i\mathcal{O}_i
\sim E^{\Delta_i-n-z}g_i^{(0)}
\sim E^{-\delta_i}g_i^{(0)}
\sim \lambda_i^{(0)}\left(\frac{E}{\Lambda_0}\right)^{-\delta_i}\,.
\end{split}
\eea
Given this expression,
\begin{itemize}
\item When $\delta_i>0$, or equivalently $\Delta_i<n+z$, this term is relevant since it grows towards the IR, in the limit $E / \Lambda_0 \rightarrow 0$.
\item When $\delta_i=0$, or equivalently $\Delta_i=n+z$, this term is marginal.
\item When $\delta_i<0$, or equivalently $\Delta_i>n+z$, this term is irrelevant since it vanishes towards the IR.
\end{itemize}

Suppose we have a non-vanishing Proca field $B_t$ which leads to a Lifshitz UV fixed point that is dual to the bulk geometry
\bea
ds^2=-e^{2u}dt^2+e^{\frac{2}{z}u}d\vec{x}^2+N_0^2du^2\,.
\eea
After identifying $r \equiv e^u$, 
\bea
ds^2=-r^2dt^2+r^{\frac{2}{z}}d\vec{x}^2+N_0^2\frac{dr^2}{r^2}\,. 
\eea
We can find that it is invariant under the isometry
\bea
t\rightarrow \lambda^z t\,,\quad
r^{-1}\rightarrow \lambda^z r^{-1}\,, \quad
\vec{x}\rightarrow \lambda \vec{x}\,,
\eea
where $r$ is not related to the energy scale anymore since its dimension is $z$. Nevertheless, it is simpler to work in these coordinates. 

Next we can introduce linear perturbations for the Proca field $\delta B_t$ around the Lifshitz fixed point, and the linear solutions near this fixed point can be written in the following form as 
\bea
\delta B_t(x^\mu,r)\sim C_t(x^\mu)~ r^{\frac{\Delta_+}{z}}+ D_t(x^\mu)~r^{\frac{\Delta_-}{z}}\,,
\eea
where we have set $\Delta_+\geq \Delta_-$ to make the first solution be the leading contribution in the UV so that $C_t$ corresponds to the external source in the standard quantization. The source in the boundary field theory is then identified as 
\bea
b_t(x^\mu)=\delta B_t(x^\mu,r)r^{-\frac{\Delta_+}{z}}=C_t(x^\mu).
\eea

Under the bulk isometry, the invariants are scalar quantities, such as $A=A_tdt\,,B=B_t dt$ and $ds^2$. Since $t\rightarrow \lambda^z t$ under scaling transformation, the isometry requires the Proca field $\delta B_t$ to transform as 
\bea
\delta B_t(x^\mu,r)\rightarrow \delta B'_t(x'^\mu,r)=\lambda^{-z}\delta B_t(x^\mu,r)
\eea
which leads to 
\bea
\begin{split}
\delta B'_t(x'^\mu,r')&=C'_t(x'^\mu)r'^{\frac{\Delta_+}{z}}+D'_t(x'^\mu)r'^{\frac{\Delta_-}{z}}\,\\
&=C'_t(x'^\mu)\lambda^{-\Delta_+}r^{\frac{\Delta_+}{z}}+
D'_t(x'^\mu)\lambda^{-\Delta_-}r^{\frac{\Delta_-}{z}}\,\\
&=\lambda^{-z}C_t(x^\mu)r^{\frac{\Delta_+}{z}}+\lambda^{-z}D_t(x^\mu)r^{\frac{\Delta_-}{z}}\,,
\end{split}
\eea
from which one concludes, 
\bea
C'_t(x'^\mu)=\lambda^{\Delta_+-z}C_t(x^\mu)\,.
\eea
Similarly, the source transforms as
\bea
b'_t(x'^\mu)=\delta B_t'(x'^\mu)r'^{-\frac{\Delta_+}{z}}=C'_t(x'^\mu)=\lambda^{\Delta_+-z}C_t(x^\mu)=\lambda^{\Delta_+-z}b_t(x^\mu).
\eea

Let us now assume that the gravitational solution with an asymptotic Lifshitz boundary has a CFT dual. Then, due to conformal symmetry, the effective action is scale-invariant under the scaling transformation, 
\bea
\int dt'd^nx'\,b'_tJ'^t=\int dtd^nx\lambda ^{n+z}\lambda^{\Delta_+-z}b_tJ'^t=\int dtd^nx \,b_tJ^t\,,
\eea
and the operator $J^t$ transforms as 
\bea
J^t\rightarrow J'^t=\lambda^{-\Delta_+-z}J^t\equiv \lambda^{-\Delta}J^t\,,
\eea
which leads to the final result for the conformal dimension $\Delta=\Delta_++n$ of the operator $J^t$.

In the main text, we perform linearized perturbations around the Lifshitz fixed points in the gravitational bulk and extract the powers in the linear solutions, in order to analyze the instabilities at linear order. This is equivalent to finding the relevant and irrelevant directions induced from operators near the fixed points, as introduced here in this Appendix.

\providecommand{\href}[2]{#2}\begingroup\raggedright\endgroup

\end{document}